\documentclass[twocolumn]{aastex701}
\usepackage{enumitem}
\usepackage{booktabs}
\usepackage{comment}
\begin{document}

\title{Resolved Maps of Gas and Dust in a Massive Quiescent Galaxy at $z=2$ from INQUEST-JWST: Evidence of Accretion and Rejuvenation}

\correspondingauthor{Sai Bhargavi Gangula}
\email{sgangula@carnegiescience.edu}

\author[0009-0005-4115-7463]{Sai Gangula}
\affiliation{Observatories of the Carnegie Institution for Science, 813 Santa Barbara Street, Pasadena, CA 91101, USA}
\affiliation{Department of Physics $\&$ Astronomy, University of Southern California, Los Angeles, CA, 90089, USA}
\email{sgangula@carnegiescience.edu}

\author[0000-0001-7769-8660]{Andrew B. Newman}
\affiliation{Observatories of the Carnegie Institution for Science, 813 Santa Barbara Street, Pasadena, CA 91101, USA}
\affiliation{Department of Physics $\&$ Astronomy, University of Southern California, Los Angeles, CA, 90089, USA}
\email{}

\author[0000-0002-4267-9344]{Meng Gu}
\affiliation{Department of Astronomy, Tsinghua University, Beijing 100084, People’s Republic of China}
\affiliation{Hong Kong Institute for Astronomy $\&$ Astrophysics, Pokfulam Road, The University of Hong Kong, Pok Fu Lam, Hong Kong}
\email{}

\author[0000-0002-5615-6018]{Sirio Belli}
\affiliation{Dipartimento di Fisica e Astronomia (DIFA), Università di Bologna, Bologna, Italy}
\email{}

\author[0000-0001-7160-3632]{Katherine E. Whitaker}
\affiliation{Department of Astronomy, University of Massachusetts, Amherst, MA 01003, USA}
\affiliation{Cosmic Dawn Center (DAWN), Denmark}
\email{}

\author[0000-0002-2784-564X]{Tania M. Barone}
\affiliation{Center for Astrophysics, Harvard $\&$ Smithsonian, Cambridge, MA 02138, USA}
\affiliation{Centre for Astrophysics and Supercomputing, Swinburne University of Technology, PO Box 218, Hawthorn, VIC 3122, Australia}
\email{}

\author[0000-0002-9861-4515]{Aliza Beverage}
\altaffiliation{NHFP Hubble Fellow}
\affiliation{Observatories of the Carnegie Institution for Science, 813 Santa Barbara Street, Pasadena, CA 91101, USA}
\affiliation{Department of Astrophysical Sciences, Princeton University, 4 Ivy Lane, Princeton, NJ 08544, USA}
\email{}

\author[0000-0001-5976-9728]{Andrea Bolamperti}
\affiliation{Max-Planck-Institut f\"ur Astrophysik, Karl-Schwarzschild-Str. 1, D-85748 Garching, Germany}
\affiliation{INAF -- IASF Milano, via A. Corti 12, I-20133 Milano, Italy}
\email{}

\author[]{Letizia Bugiani}
\affiliation{Dipartimento di Fisica e Astronomia (DIFA), Università di Bologna, Bologna, Italy}
\affiliation{INAF, Osservatorio di Astrofisica e Scienza dello Spazio, Bologna, Italy}
\email{}

\author[0000-0001-7782-7071]{Richard S. Ellis}
\affiliation{University College London, Department of Physics $\&$ Astronomy, Gower Street, London WC1E 6BT, UK}
\email{}

\author[0000-0002-7613-9872]{Mariska Kriek}
\affiliation{Leiden Observatory, Leiden University, P.O. Box 9513, 2300 RA Leiden, The Netherlands}
\email{}

\author[0000-0002-6479-6242]{Allison Matthews}
\affiliation{Observatories of the Carnegie Institution for Science, 813 Santa Barbara Street, Pasadena, CA 91101, USA}
\email{}

\author[0000-0003-2804-0648]{Themiya Nanayakkara}
\affiliation{Centre for Astrophysics and Supercomputing, Swinburne University of Technology, PO Box 218, Hawthorn, VIC 3122, Australia}
\email{}


\begin{abstract} 
Quiescent galaxies in the distant universe exhibit a range of gas content that may indicate a variety of quenching processes are at play. Mapping the distribution and kinematics of the gas can illuminate its origins, but nearly all such observations have been unresolved. We present JWST/NIRSpec IFU observations of MRG-M0138, a gravitationally lensed, massive quiescent galaxy at $z\sim2$ observed as part of the INQUEST-JWST survey. We use \ion{Na}{1}~D absorption, which we detect in excess of the stellar absorption over most of the galaxy, to trace the kinematics and spatial distribution of the neutral gas in 219 spatial bins. The gas exhibits clear rotation that is kinematically aligned with the stellar disk. Both the gas and dust have a complex spatial structure, including an off-nuclear clump, a dust lane, and patches in the outer disk. The non-equilibrium distribution suggests that the gas was accreted. Analysis of the galaxy's star formation history supports this interpretation by indicating a rejuvenation event $500$ Myrs ago. We identify two plausibly associated galaxies and suggest that tidal interactions are a likely source of the accreted gas. Our results indicate that some of the variation in gas content among early quiescent galaxies is not related to differences in gas consumption timescales. The detection of a gas clump at a projected distance of $\sim90$~pc from the known supermassive black hole illustrates a mechanism to fuel the episodic AGN feedback that may maintain quiescence.
\end{abstract} 

\section{Introduction} 
\label{sec:intro}
A key open question in galaxy evolution is whether quenching at high redshifts proceeds primarily through gas consumption, gas removal via outflows, gas heating through feedback, or internal processes that suppress star formation without consuming gas \citep{Man18}. Addressing this question requires robust measurements of the cold gas content in galaxies as a function of the time since quenching. Spatially resolved observations of the gas distribution and kinematics are especially valuable to determine the origin of any remaining interstellar medium (ISM; including incomplete consumption, stellar mass loss, or external accretion) and to detect and characterize outflows, which provide crucial diagnostics of feedback processes and their role in quenching.

Observations of CO emission have demonstrated that quiescent galaxies do not form a homogeneous population in their molecular gas content. For example, CO$(2-1)$ emission has been detected in roughly half of quiescent and post-starburst galaxies at $z\sim0.6-0.7$, and the presence of molecular gas has been linked either to enhanced late-time star formation (rejuvenation; \citealt{woodrum22}) or to quenching that occurs prior to gas removal or heating \citep{bezanson22, setton25}. In contrast, several studies report non-detections of CO that place stringent upper limits on the molecular content of $z \sim 0.7-1.5$ quiescent galaxies, pointing to more efficient gas removal or depletion \citep{sargent15, spilker18, rachel19, williams21, suess25}. CO observations are a direct but observationally expensive probe of molecular gas in distant quiescent galaxies, and even when detections are achieved, they are typically spatially unresolved. Furthermore, the conversion between the CO luminosity and the molecular gas mass is uncertain \citep{bolatto13}.

An alternative approach is to use dust emission at (sub-)millimeter wavelength as an indirect tracer of total molecular gas mass, assuming a gas-to-dust ratio, and this technique has been widely applied across cosmic time \citep{scoville16}. Stacking analyses have revealed substantial dust reservoirs in quiescent galaxies out to $z\sim2$, suggesting that a significant mass of gas was not efficiently consumed or ejected during quenching \citep{gobat18}. On the other hand, deep ALMA observations of individual high-redshift quiescent galaxies showed a wide range of dust emission and inferred gas and dust mass \citep{hayashi18, whitaker21, morishita22, lee24, spilker25}. Although dust continuum emission provides a rapid and efficient means of estimating ISM masses, the derived gas masses are sensitive to assumptions about dust temperature and gas-to-dust ratio, leading to potentially large systematic uncertainties \citep{madden14, magnelli14, privon18, liang19, whitaker21, gobat22, spilker25}. It remains unclear whether the coexistence of CO and dust detections and non-detections across redshift reflects a diversity in gas depletion or removal timescales among quiescent galaxies, or variations in dust temperature or the gas-to-dust ratio \citep{whitaker21_1, lorenzon25b}.

Cool and warm gas in the ISM and associated outflows can be effectively traced using absorption and emission lines in the rest--frame ultraviolet (UV). In distant star-forming galaxies, these gas absorption  features are almost ubiquitously detected and are typically blueshifted with respect to the systemic velocity, providing direct evidence for large scale, star-formation driven outflows \citep[e.g.,][]{shapley03, steidel04, weiner09, rubin10, erb12, kornei12}. In distant quiescent galaxies, however, these absorption lines are challenging to detect against the faint UV continuum, with detections mainly confined to post-starburst systems at lower redshifts \citep{maltby19, tremonti07, taylor24} and a few $z \gtrsim 4$ galaxies recently observed with JWST \citep{valentino25,wu25}.

At optical wavelengths, low-ionization absorption lines, such as \ion{Na}{1}~D and \ion{Ca}{2} H $\&$ K, provide complementary probes. In particular, \ion{Na}{1} D is a sensitive tracer of neutral hydrogen due to its low first ionization potential (5.14 eV), and Na-traced outflows have been extensively observed in the local universe \citep{heckman00, rupke02, rupke05, rupke05_1, martin05, martin06}. At higher redshifts, these rest-frame optical transitions are redshifted into the near-infrared, where pre-JWST observations were severely limited. The high sensitivity and spectral resolution of JWST now overcome many of these challenges.

JWST observations have begun to directly reveal neutral gas in high-redshift quiescent and post-starburst galaxies. At $z \sim 1-3$, outflows have been traced using the \ion{Na}{1}~D and \ion{Ca}{2}~H and K lines \citep[e.g.,][]{belli24,liboni25,sun25}, while the \ion{Mg}{2} and \ion{Fe}{2} UV lines have now been observed in a few $z \gtrsim 4$ systems \citep{valentino25,wu25}. In many cases, the inferred mass outflow rates greatly exceed the ongoing star formation rates, implying that such outflows are capable of rapidly suppressing star formation and are likely driven by recent or ongoing AGN feedback \citep{belli24, eugenio24, davies24, park24, taylor24, taylor26}. Other early-quenched galaxies have retained neutral gas traced by \ion{Na}{1}~D absorption, but without the blueshift characteristic of large-scale outflows \citep{davies24,park24}.

Taken together, observations of dust continuum, CO, and \ion{Na}{1}~D suggest a possible range of gas and dust masses in $z \gtrsim 1.5$ quiescent galaxies. Although gas may at times be inflowing or outflowing, in other cases, it appears to be an \emph{in situ} ISM. The combination of significant gas reservoirs and low star-formation rates would be an important indication of a mechanism that reduces star-formation efficiency, such as stabilization by a bulge \citep{martig09}, thereby enabling the gas and dust to survive for several Gyr after quenching \citep{gobat18,lee24}. However, to interpret gas reservoirs as evidence of quenching mechanisms, we must first understand not just the mass of gas, but also its origin. Although gas in early quiescent galaxies is often interpreted as residual material that was not consumed or removed when the galaxy quenched, it could instead have been replenished later by stellar mass loss or accretion, perhaps associated with rejuvenation in star formation. Indeed, in the local universe, accretion and subsequent grain growth is thought to be responsible for the dust observed in about half of early-type galaxies \citep{Martini13}. Spatially resolved observations played a central role in that inference, because they often revealed a disturbed morphology. By contrast, little is known about the spatial distribution and resolved kinematics of gas in high-redshift quiescent galaxies. To date, the \ion{Na}{1}~D absorption has been spatially resolved in only one such galaxy \citep{eugenio24}, where evidence for neutral gas outflows driven by AGN feedback was found in a $z \sim 3$ quiescent galaxy. 

In this work, we use gravitational lensing to spatially resolve the neutral gas distribution and kinematics, as well as the dust attenuation, in MRG-M0138, a massive quiescent galaxy at $z = 1.95$ \citep{newman18}. Situated behind the galaxy cluster MACS~J0138.0–2155, MRG-M0138 is extraordinarily bright compared to all other known high-redshift quiescent galaxies. The combination of a high stellar mass with an areal magnification of $\mu = 29^{+13}_{-11}$ \citep{suyu25,ertl25,newman_bh} makes MRG-M0138 a singular laboratory for dissecting an early quiescent galaxy at a remarkable level of detail. We use JWST/NIRSpec IFU spectroscopy to map the distribution and kinematics of gas in MRG-M0138, focusing mainly on \ion{Na}{1}~D absorption, to explore the origin of the gas reservoir and its possible connection to quenching. MRG-M0138 is one of five gravitationally lensed quiescent galaxies observed as part of the INQUEST-JWST survey (INvestigating QUEnching in Strongly lensed Targets with JWST; GO-2345 and GO-4903).

This paper is organized as follows. In Section \ref{sec:obs}, we describe the observations and data. Section \ref{sec:gas_absp} details the analysis of the \ion{Na}{1}~D absorption lines. The spatially resolved maps and gas mass estimates are presented in Section \ref{sec:gas_dust}. In Section \ref{sec:sed}, we describe the SED modeling used to infer the galaxy’s star-formation history. We introduce candidate companion galaxies to MRG-M0138 in Section \ref{sec:companion}. The origin of the gas is discussed in Section \ref{sec:discussion}. Finally, we summarize our results in Section \ref{sec:conclusion}. Where required, we adopted a \citet{kroupa01} initial mass function (IMF) and a flat $\Lambda$CDM cosmology with $\Omega_m$ = 0.31 and $\rm H_0=67.4~km~s^{-1}~Mpc^{-1}$ \citep{planck18}.  

\begin{figure*}[t]
    \centering
    \includegraphics[width=7in]{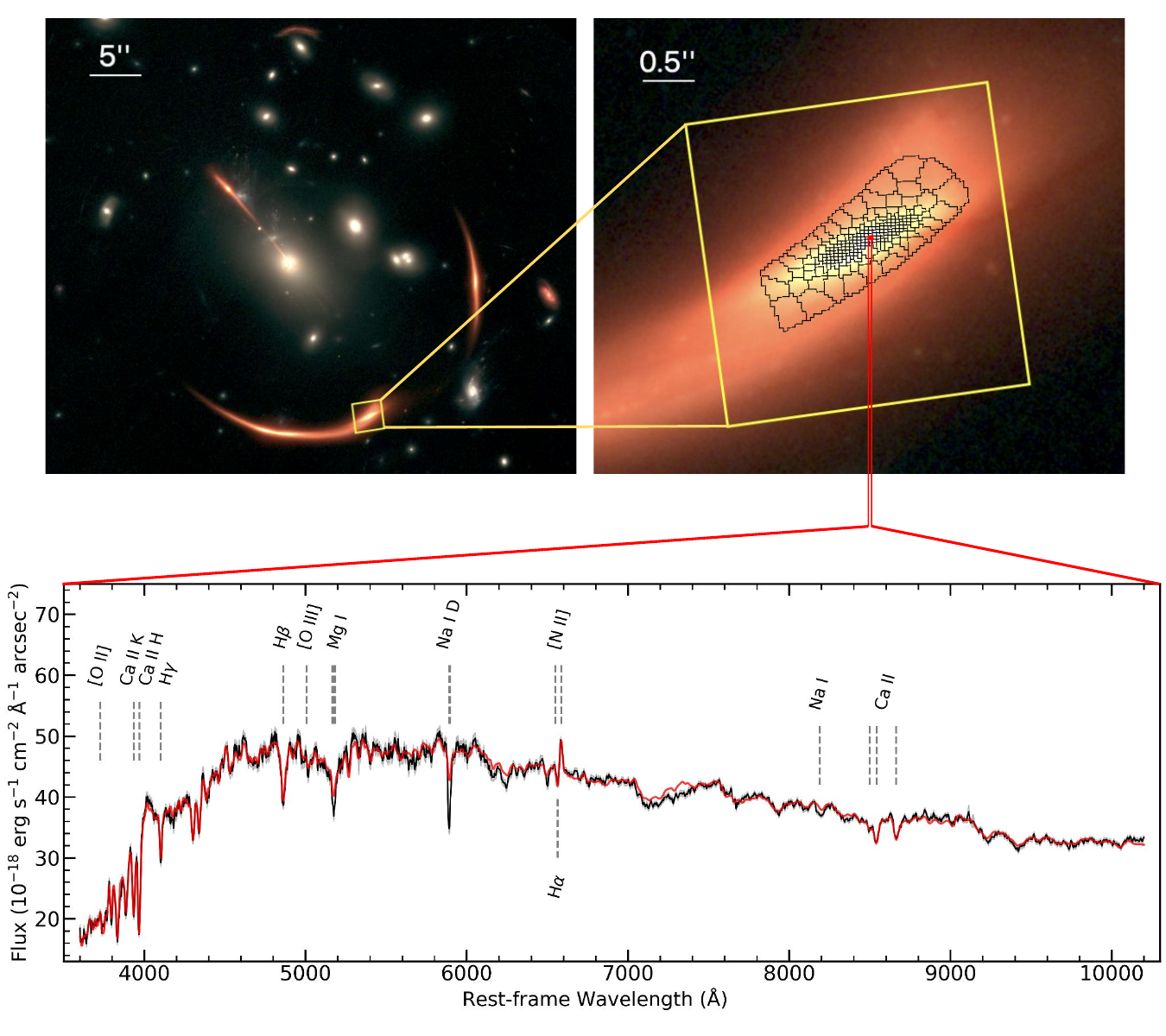}
    \caption{JWST/NIRSpec observations of the lensed quiescent galaxy MRG-M0138 at $z = 1.948$. \textbf{Top left:} JWST/NIRCam image of the lensing galaxy cluster MACS J0138.0--2155 obtained with the F115W, F150W, and F356W filters. MRG-M0138 appears as a bright, red, multiply imaged source. The yellow box indicates the NIRSpec IFU field of view. \textbf{Top right:} Zoomed-in view centered on the IFU field of view. The overlaid black grid delineates the spatial bins (219 in total). \textbf{Bottom:} The spectrum extracted from one example spatial bin (black), with the $1\sigma$ uncertainty shown as the grey shaded region. The red curve represents the bestfit stellar continuum and nebular emission model derived using \texttt{pPXF}. Prominent spectral features, including \ion{Na}{1}~D and \ion{Ca}{2} H $\&$ K, are indicated.}
    \label{fig:obs}
\end{figure*}

\section{Target and Observations}
\label{sec:obs}

MRG-M0138 has a total stellar mass of $\rm M_{\star} = 2.2 ^{+1.4} _{-0.7} \times 10^{11} M_\odot$ \citep{newman_bh}, based on stellar population synthesis (SPS) modeling. It contains a bulge and a disk component, with the disk contributing 62\% of the F200W light. We employed the \texttt{iso\_halo} lens model of \citet{ertl25} to reconstruct the galaxy in the source plane. The effective radius of the entire system, defined as the semi-major axis of the ellipse containing half of the light, is $\rm R_e = 2.7^{+0.8} _{-0.5}~kpc$, while the bulge effective radius is $\rm R_{e,bulge} = 0.8 ^{+0.2} _{-0.1} ~kpc$ \citep{newman_bh}. MRG-M0138 is also one of the two quiescent galaxies with ALMA dust continuum detections reported by \citet{whitaker21}. Its 1.3~mm emission was centrally located and unresolved by the $\sim 1\farcs3$ beam.

We used the same JWST and HST observations of MRG-M0138 that were described in detail by~\citet{newman_bh}. These include images from JWST/NIRCam obtained through programs GO-2345 and DD-6549, which provided deep imaging in six filters (F115W, F150W, F200W, F277W, F356W, and F444W), along with an 87-minute HST/ACS F555W exposure. Integral field spectroscopy of one image of MRG-M0138 (the image enclosed in the yellow box in the upper panel of Figure~\ref{fig:obs}) was obtained using JWST/NIRSpec with the G140M/F100LP (137~min exposure) and G235M/F170LP (126~min) grating/filter pairs.

The data reduction and processing steps, including background subtraction and modeling of spectral ``wiggles'' caused by the undersampled NIRSpec point spread function (PSF), were described by \citet{newman_bh}. Voxels (cells of the data cube) in the resulting data cube span $0\farcs05$ on sky and 132 km~$\text{s}^{-1}$ in wavelength (logarithmically binned). When fitting the full spectrum to model the starlight, we used a data cube built by splicing together the G140M and G235M data cubes at $\lambda_{\rm rest} = 0.615$ $\mu$m. When fitting the \ion{Na}{1}~D line, which is covered by both the G140M and G235M observations, we used both data sets individually.

We spatially binned the spectra to a target signal-to-noise ratio (SNR) of 40, using the {\tt vorbin} algorithm  \citep{vorbin,Cappellari03} to define 219 spatial bins, as in \citet{newman_bh}. The spatial resolution reaches $\sigma_{\rm psf} = 11~{\rm mas} = 91$~pc (source plane) at the galaxy center, and it degrades toward the galaxy outskirts where larger bins are needed to reduce noise. The error spectra were rescaled from the pipeline-provided values based on residuals to the stellar continuum model fits (Section~\ref{sec:continuum}). We also measured PSF-matched photometry in each spatial bin. For full details of these steps, we refer the reader to \citet{newman_bh}.

\section{Measuring Gas in Absorption}
\label{sec:gas_absp}

\begin{figure*}[t]
    \centering
    \includegraphics[width=7in]{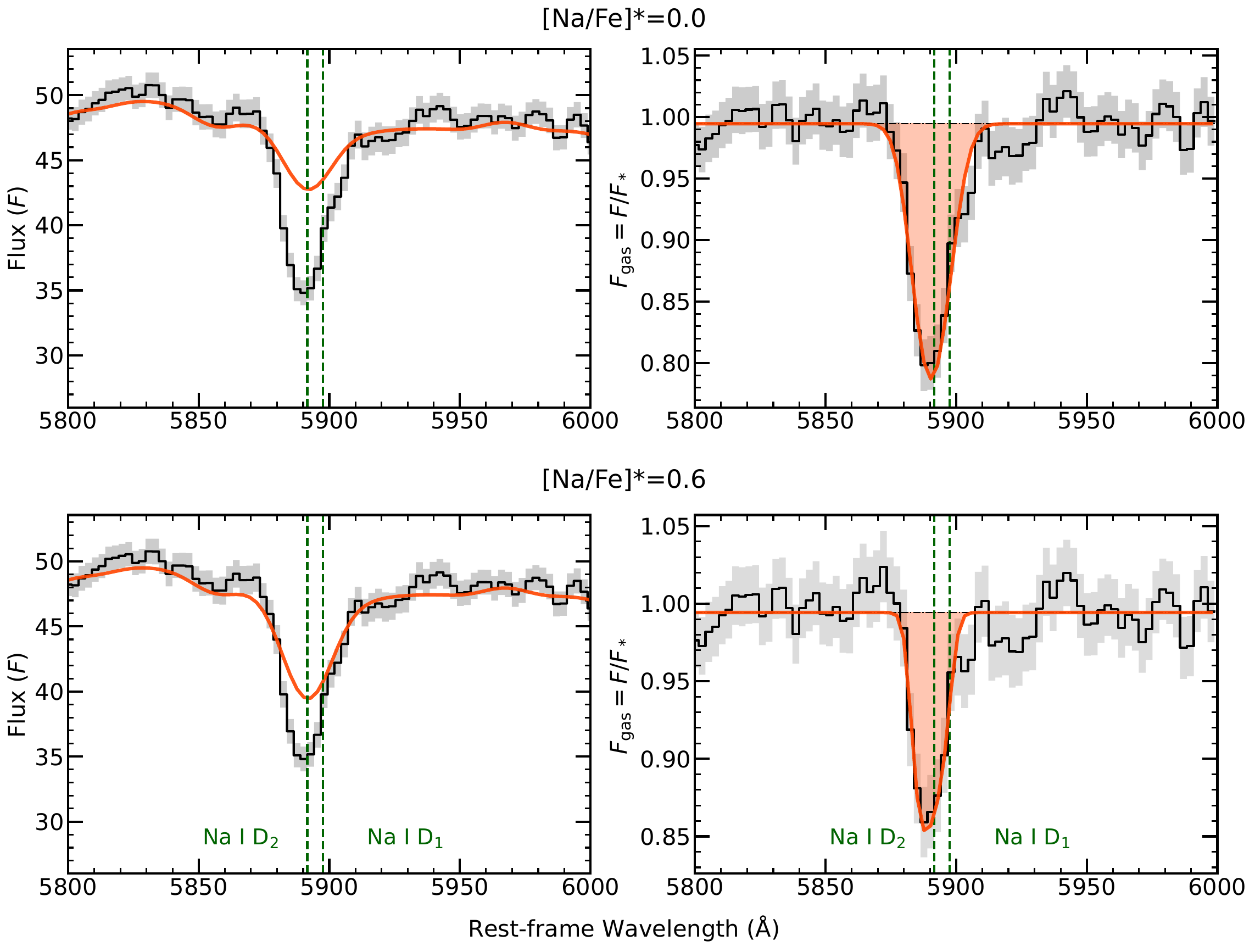}
    \caption{\ion{Na}{1}~D absorption lines observed in an example spatial bin (same bin used in Figure~\ref{fig:obs}) in the G140M spectra. \textbf{Left panels:} The observed spectrum $F$ (black) is compared to the best-fit \texttt{pPXF} stellar model (orange) with [Na/Fe]* = 0 in the top panel and [Na/Fe]* = +0.6 in the bottom panel. Wavelengths are shifted to the systemic rest-frame of the galaxy. \textbf{Right panels:} The corresponding gas absorption spectra $F_{\rm gas} = F / F_*$ (black) and the best-fit Voigt model (orange) with [Na/Fe]* = 0 in the top panel and [Na/Fe]* = +0.6 in the bottom panel. In all panels, the green dashed lines show the rest-frame wavelengths of the \ion{Na}{1} D doublet lines.}
    \label{fig:na}
\end{figure*}

\subsection{Modeling the Stellar Continuum}
\label{sec:continuum}
To robustly identify absorption produced by neutral gas, it is imperative to accurately remove the stellar absorption contribution. We modeled the stellar continuum using the \texttt{pPXF} \citep{ppxf23} code with simple stellar population (SSP) templates from the E-MILES library \citep{emiles}. Templates were included for several nebular emission lines. Dust was modeled using the attenuation curve of \citet{calzetti94}. Several wavelength regions were masked, including those affected by potential gas absorption or strong sensitivity to the IMF or $\alpha$-enhancement. We refer the reader to~\citet{newman_bh} for a more detailed description of the spectral fitting procedure. The SPS modeling was performed on a spliced data cube in which the \ion{Na}{1}~D line is covered by the G140M data. When we modeled the \ion{Na}{1}~D absorption in the G235M data, we modified the stellar continuum model to account for the different instrumental resolution. Figure~\ref{fig:obs} shows the spectrum and the \texttt{pPXF} best-fit model in an example spatial bin. 

The strengths of the \ion{Na}{1}~D and \ion{Ca}{2} H and K absorption features depend on the age and metallicity of the stellar population, both of which are constrained by our modeling. But they also depend on the elemental abundance ratios [Na/Fe]* and [Ca/Fe]*.\footnote{[Na/Fe]* and [Ca/Fe]* are the stellar abundance ratios.} While [Ca/Fe]* shows little variation in local early--type galaxies \citep{jonas12,conroy14,krause21}, numerous studies find that [Na/Fe]* is significantly super-solar in these systems \citep{la17,parikh21,meng22,parikh24}, suggesting a similar enhancement may hold for MRG-M0138. To explore the impact of Na enhancement on our results, we used the response functions from the {\tt alf} spectral fitting code \citep{alf23}, which quantify how the spectrum of a given age and metallicity changes in response to variations in [Na/Fe]*.

We focused on a sodium enhancement of $[{\rm Na/Fe}]^* = +0.6$~dex, which is approximately the maximum observed in low-redshift early-type galaxies (see references above), to evaluate the maximum expected effect of Na enhancement. The response functions are computed for SSPs, and we matched them to the observations using the mean light-weighted age and metallicity from our {\tt pPXF} fits in each spatial bin. We performed bilinear interpolation across the response function grid in log(age) and metallicity. The resulting response function was then convolved by $\sigma = \sqrt{\sigma_{\rm inst}^2 + \sigma_{*}^2 - \sigma_{\rm rf}^2}$, where $\sigma_{\rm inst}$ is the instrumental resolution\footnote{The instrumental resolution at \ion{Na}{1}~D is 102 $\rm km~s^{-1}$ for the G140M data and 173 $\rm km~s^{-1}$ for the G235M data.}, $\sigma_*$ is the local stellar velocity dispersion, and $\sigma_{\rm rf} = 100$~km~s${}^{-1}$ is the response function resolution. The broadened response function was subsequently shifted to the stellar velocity in that bin. Finally, the baseline \texttt{pPXF} fit, which we take to have [Na/Fe]* = 0, was multiplied by the modified response function to produce the sodium-enhanced model spectrum with [Na/Fe]* = +0.6.

\begin{figure*}[t]
    \centering
    \includegraphics[width=6.75in]{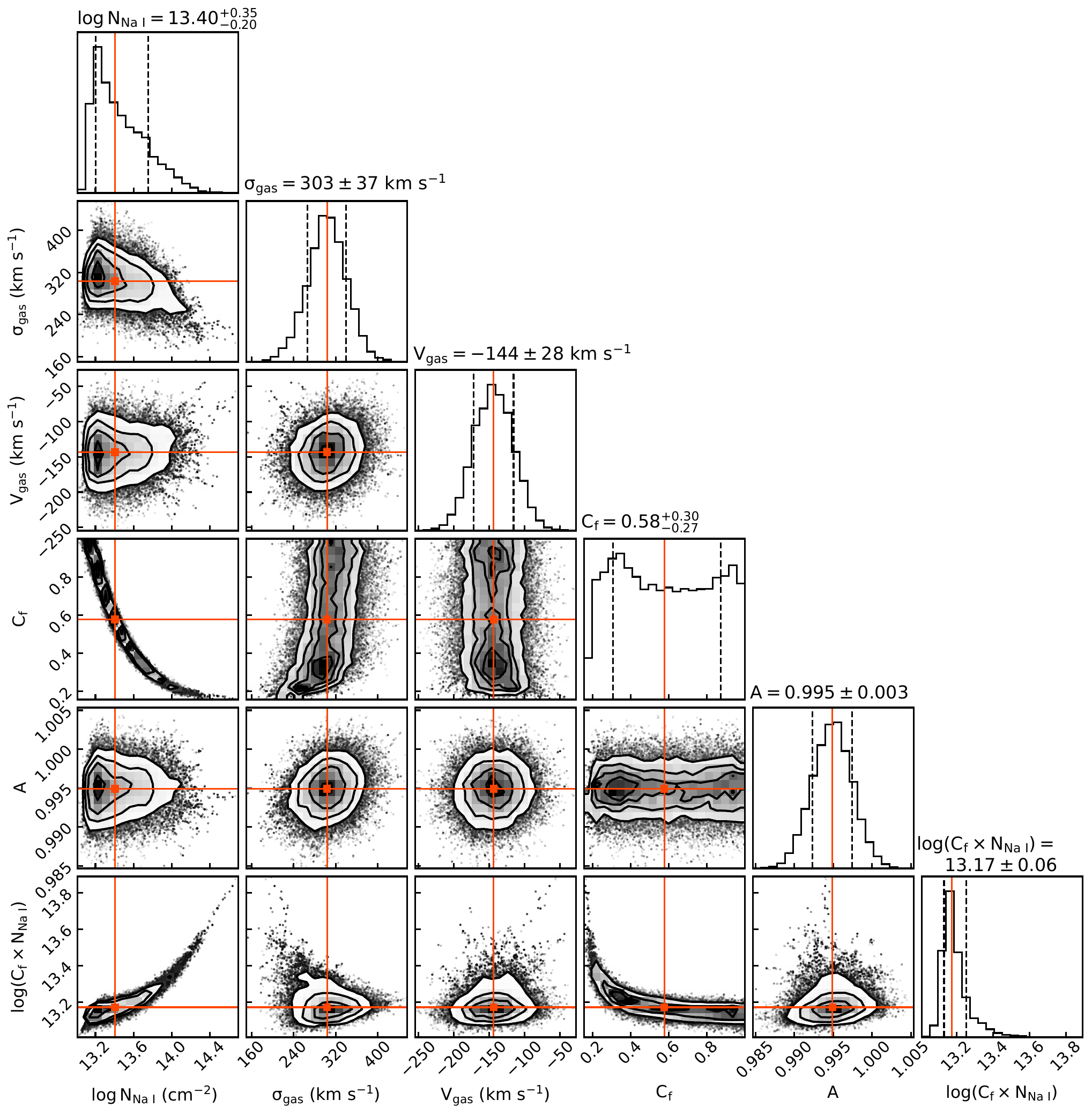}
    \caption{Corner plot of the posterior distributions for the Partial Covering fraction Voigt profile parameters characterizing excess \ion{Na}{1}~D absorption, obtained after dividing the observed spectrum by the best-fit stellar continuum \texttt{pPXF} model, in the same spatial bin used in the Figure~\ref{fig:obs} and \ref{fig:na}. For each parameter, the orange line marks the median, and the black dashed lines indicate the $16^{\text{th}}$ and $84^{\text{th}}$ percentiles. In addition to the model parameters, we include an extra row and column for the derived quantity $\log( N \times C_f)$. Although not a free parameter in the fit, this combination is more tightly constrained than either $\log N_{\rm Na~I}$ or $C_f$ individually, highlighting a key degeneracy in the model.}
    \label{fig:corner_}
\end{figure*}

\subsection{Voigt Profile Fitting}
\label{sec:voigt}
We divided the stellar continuum model $F_*$ from the observed spectrum $F$ to isolate the excess absorption due to neutral gas, $F_{\rm gas}$.  In the discussion below, we focus on the \ion{Na}{1}~D doublet for concreteness, but in Section~\ref{sec:ca} we will apply the same technique to the \ion{Ca}{2} H $\&$ K doublet. We modeled $F_{\rm gas}$ using a standard partial covering fraction model \citep{rupke05}:
\begin{equation}
\ F_{\rm gas} = \frac{F} {F_*} = A(1 - C_{f} + C_f~ e^{-\tau_b (\lambda)-\tau_r (\lambda)}).
\label{eqn:Fgasmodel}
\end{equation}
Here, $C_f$ is the covering fraction of the absorbing gas against the background continuum source, and $\tau_b (\nu)$ and $\tau_r (\nu)$ are the optical depth profiles of the blue (\ion{Na}{1}~D $\lambda$ 5891.58~\AA) and red (\ion{Na}{1}~D $\lambda$ 5897.55~\AA) doublet lines, respectively. $A$ is a constant that accounts for small, local mismatches between the continuum normalizations of the observation and the stellar model.

We assume that the optical depth $\tau(\lambda)$ has a Voigt profile characterized by a column density $N_{\rm Na~I}$, a velocity of the gas ($V_{\rm gas}$) relative to the galaxy systemic redshift $z = 1.948$, and a Doppler parameter $ b = \sqrt{2} \sigma_{\rm gas}$, where $\sigma_{\rm gas}$ is the velocity dispersion of the gas. Calculation of the Voigt profile requires the wavelength, oscillator strength $f_{\rm osc}$, and damping coefficient $\Gamma$ for each of the doublet components, which share a common $N$, $V_{\rm gas}$, and $\sigma_{\rm gas}$. We used values $ \{\Gamma_{\rm b}, f_{\rm osc,b},\lambda_{\rm b}\}$ = $\{61.542 \times 10^6 \ \text{s}^{-1},0.64050,5891.583$ \AA $\}$ and $\rm \{\Gamma_r,f_{\rm osc,r},\lambda_{r}\}$ = $\{61.353 \times 10^6 \ \text{s}^{-1}, 0.31992, 5897.558$ \AA $\}$ for the blue and red doublet lines, respectively \citep{mortonIII}. All wavelengths are in vacuum.  When we fitted the data, the $F_{\rm gas}$ model in equation~\ref{eqn:Fgasmodel} was convolved by the instrumental resolution. Therefore $\sigma_{\rm gas}$ is the intrinsic gas velocity dispersion. We note that a Gaussian optical depth profile is often used in the literature \citep[e.g.,][]{murga15,davies24}. Since the Gaussian and Voigt profiles differ only when the absorption is strongly damped, which does not occur for Na absorption, this distinction should not be important. 

\begin{figure*}
  \centering
  \includegraphics[width=7in]{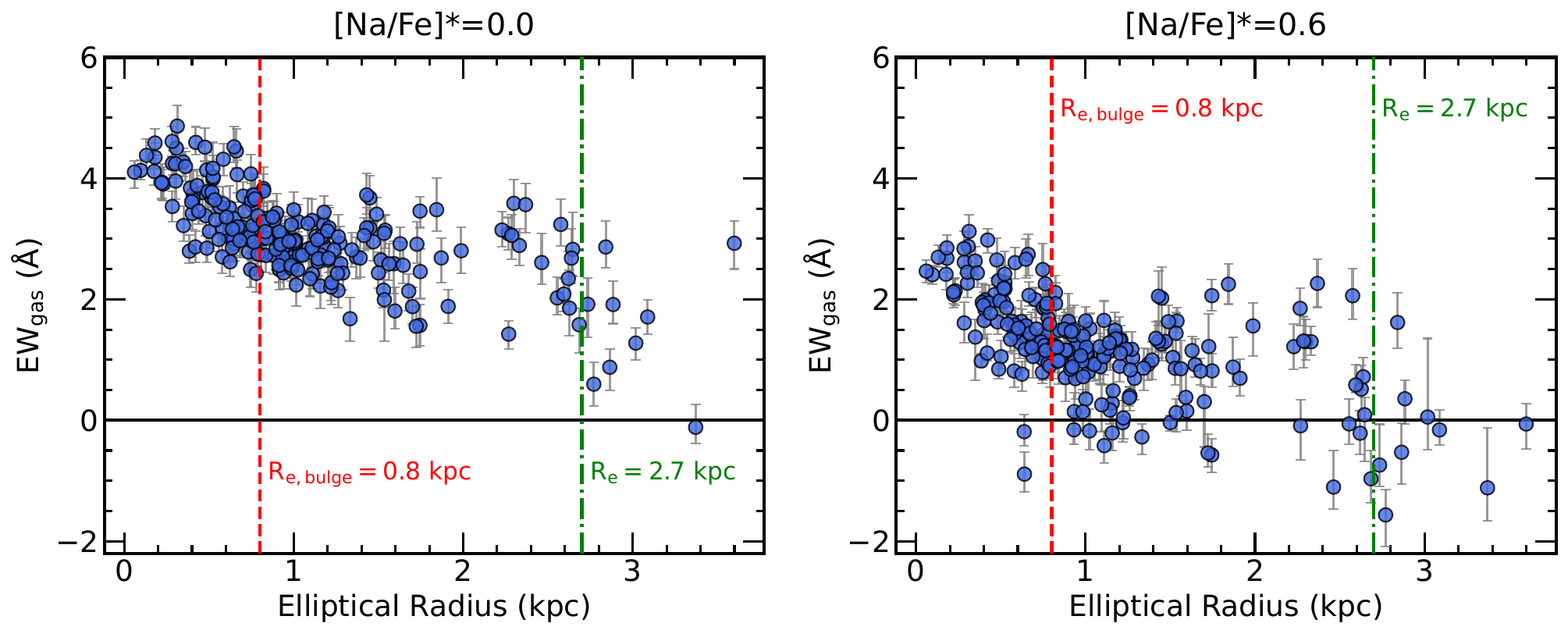}
  \caption{Radial profiles of the gaseous \ion{Na}{1}~D equivalent width ($\rm EW_{gas}$), measured from the residual absorption after dividing the observed spectrum by the best-fit stellar continuum model. The elliptical radius is defined as $(x^2 + (y/q)^2)^{1/2}$, where $x$ and $y$ are projections of a bin onto the major and minor axes of the galaxy, respectively, and $q = 0.28$ is the mean axial ratio of the starlight in the F200W filter. Radial profiles are shown for two assumed stellar sodium abundances: [Na/Fe]*$=0.0$ (\textbf{left panel}) and [Na/Fe]*$=+0.6$ (\textbf{right panel}). We averaged the $\rm EW_{gas}$ measurements from the G140M and G235M observations. In both plots, the dashed red line denotes the bulge effective radius $\rm (R_{e, bulge})$, and the dot dashed green line denotes the effective radius of the entire galaxy $\rm (R_{e})$.}
  \label{fig:ew_radial}
\end{figure*}

In summary, the model describing non-stellar sodium absorption is characterized by five parameters: $\log N_{\rm Na~I}$, $C_f$, $V_{\rm gas}$, $b$, and $A$. In each spatial bin, we fit such a model to the $F_{\rm gas}$ spectrum over the wavelength region 5800--6000~\AA. To obtain accurate constraints on the parameter uncertainties and degeneracies, we used \texttt{emcee} \citep{emcee13}, an affine-invariant Markov Chain Monte Carlo (MCMC) \citep{hogg18} ensemble sampler. We adopted broad, uniform priors for all the parameters. 

\subsection{Detection of Interstellar Sodium Absorption}
\label{sec:namodels}
The \ion{Na}{1} D feature is covered by both the G140M and G235M gratings. To analyze the \ion{Na}{1} D absorption, we fit four combinations of data and models: we used either the G140M or the G235M data, and an assumed value of [Na/Fe]* = 0 or +0.6 for the stellar continuum models. This approach allowed us to assess the robustness of our results. Figure~\ref{fig:na} illustrates the spectrum, stellar continuum model, residual gas absorption, and the model fits for one example spatial bin. 

Figure~\ref{fig:corner_} shows the resulting parameter constraints. As expected, a prominent degeneracy is observed between the column density $N_{\rm Na~I}$ and the covering fraction $C_f$, while their product $C_f \times N_{\rm Na~I}$ is more tightly constrained than either parameter individually. This reflects the fact that if the absorption is not saturated, the line strength depends on the total number of absorbing atoms within the spatial resolution element. Although the degeneracy can be broken by measuring the two doublet components individually, the lines are blended in MRG-M0138 by the high velocity dispersion. In most spatial bins, $C_f$ is very weakly constrained, with the data mainly providing a lower limit to ensure sufficient absorption depth. We also note a secondary covariance between $C_f$ and $\sigma_{\rm gas}$, which we will discuss in Section~\ref{sec:gaskin}. 

We examined the significance with which excess (gaseous) \ion{Na}{1} absorption is detected by performing a model comparison using the Akaike Information Criterion (AIC), which quantifies the relative quality of models by balancing goodness-of-fit against model complexity. It is defined as ${\rm AIC} = 2k - 2 \ln L$, where $k$ is the number of parameters and $L$ is the maximum likelihood. Specifically, we compared a Voigt profile model, representing a physical absorption line, to a simpler second-order polynomial model that does not include any line component. Higher values of the difference $\Delta {\rm AIC} = {\rm AIC}_{\rm Polynomial} - {\rm AIC}_{\rm Voigt}$ indicate stronger support for the absorption line model. When using the stellar models with ${\rm [Na/Fe]^*} = 0$, we find strong evidence for gas absorption in virtually all spatial bins. Specifically, we found $\Delta {\rm AIC} > 10$ in 97\% (G140M) or 99\% (G235M) of the 219 bins. When assuming ${\rm [Na/Fe]^*} = +0.6$, we find strong evidence in 64\% (G140M) or 76\% (G235M) of spatial bins. 

\begin{figure*}[t]
    \centering
    \includegraphics[width=7in]{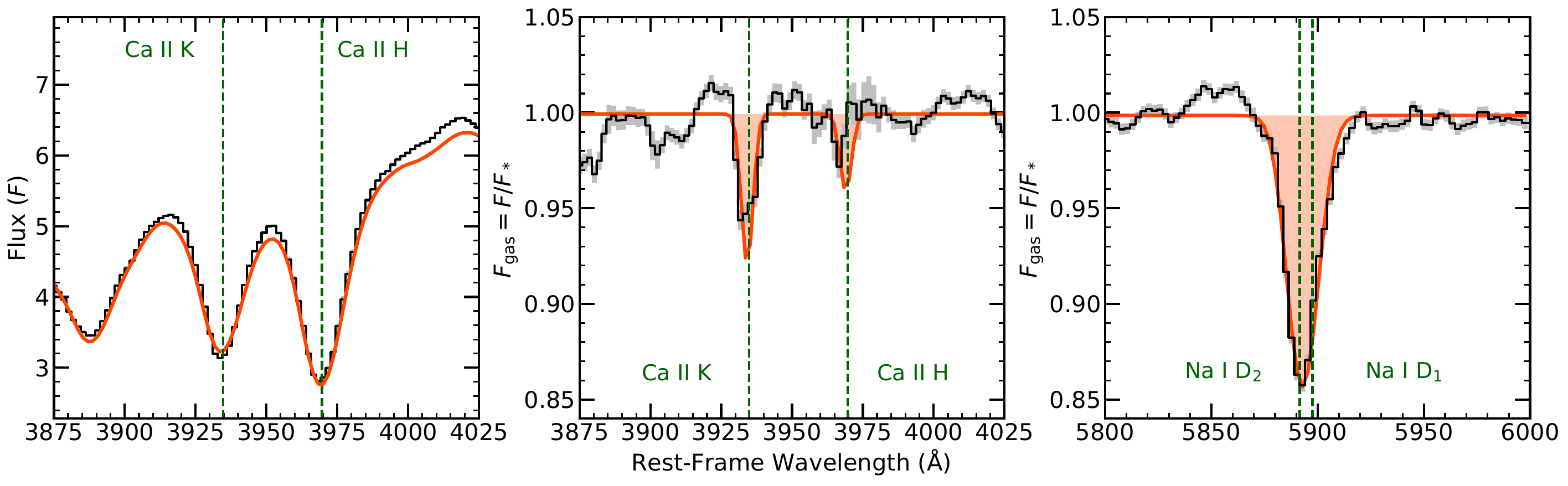}
    \caption{Detection of interstellar \ion{Ca}{2} H, K absorption in the spatially integrated spectrum. \textbf{Left panel}: The observed stacked spectrum (black) is compared to the best-fit \texttt{pPXF} stellar model (orange). \textbf{Middle panel}: The ratio of the observed spectrum to the best-fit stellar model (black) shows evidence of excess absorption at \ion{Ca}{2}~H, K, which is modelled (orange) as discussed in Section~\ref{sec:ca}. \textbf{Right panel}: As in the middle panel, but in the wavelength region around \ion{Na}{1}~D. The sodium absorption is much stronger, explaining why it is detectable in individual spatial bins.}
    \label{fig:ca}
\end{figure*}

This demonstrates that gaseous \ion{Na}{1} absorption is widespread. We visualize this in Figure~\ref{fig:ew_radial}, which shows radial profiles of the equivalent width of the gas ($\rm EW_{gas}$). For this figure, we used a different set of models in which $F_{\rm gas}$ was fit with a Gaussian profile. Although this simple model does not account for the doublet nature of the \ion{Na}{1}~D absorption, the components are highly blended in our data, and we find that the Gaussian models provide an acceptable fit. The utility of a Gaussian model here is that it can allow for negative values of $\rm EW_{gas}$, which allows us to better visualize the detection significance. We averaged the $\rm EW_{gas}$ values obtained from the G140M and G235M data using inverse variance weights, and based on the scatter between the two measures, we increased the formal uncertainties by a factor of 2.

Consistent with the AIC analysis, Figure~\ref{fig:ew_radial} shows strong evidence of excess \ion{Na}{1}~D absorption in virtually all bins if ${\rm [Na/Fe]^*} = 0$ and in a majority if ${\rm [Na/Fe]^*} = +0.6$. This absorption peaks in the central regions corresponding to the effective radius of the bulge (red dashed line), while our measurements extend to roughly the effective radius of the galaxy as a whole (green dot dashed line; Section~\ref{sec:obs}). It is evident that [Na/Fe]* = 0.6 is probably too high at elliptical radii $\gtrsim$ 2 kpc, where many gas absorption measurements would fall significantly below zero equivalent width, although such a sodium enhancement may be possible in the central regions. At larger radii, the gas absorption exhibits a wide range of EW; as we will show in Section~\ref{sec:gas_dust}, the gas and dust in the outer galaxy are patchy. While we robustly detect excess absorption in some of the outer bins, its presence in other cases depends on the assumed ${\rm [Na/Fe]^*}$. 

\subsection{Calcium Absorption}
\label{sec:ca}
Having detected widespread \ion{Na}{1}~D absorption, we examined the spectra for signs of \ion{Ca}{2}~H $\&$ K absorption. Unlike \ion{Na}{1}~D, visual inspection did not show convincing evidence of \ion{Ca}{2}~H $\&$ K absorption in individual spatial bins. We generally found that the amplitude of possible \ion{Ca}{2}~H $\&$ K absorption was comparable to the correlated residuals from the {\tt pPXF} fits. This suggests that any further modeling would be limited by systematics.

Therefore, we did not attempt to map interstellar \ion{Ca}{2} absorption. However, to improve sensitivity to weak \ion{Ca}{2} absorption, we constructed a stacked spectrum representing the full galaxy. The observed spectra in all spatial bins were averaged, with weights equal to the source-plane area of the bin, and the same averaging was applied to the stellar continuum models. This stacked spectrum shows evidence of gaseous \ion{Ca}{2} absorption, as seen in Figure~\ref{fig:ca}. The \ion{Na}{1}~D absorption in the same composite spectrum (Figure~\ref{fig:ca}, right panel) is much stronger, explaining why it is detectable in the individual spatial bins.

Formally, the detection of \ion{Ca}{2} H $\&$ K absorption is highly significant, with $\Delta {\rm AIC} = 214$ following Section~\ref{sec:namodels}. However, given the very low random noise in the stacked spectrum, the residuals are dominated by systematic uncertainties in the SPS model fits rather than statistical noise. The depth of the \ion{Ca}{2} absorption is only about a factor of two larger than the amplitude of nearby fit residuals (Figure~\ref{fig:ca}, middle panel), and we therefore consider the detection of gaseous \ion{Ca}{2} absorption to be likely, though not definitive. We proceed with modeling \ion{Ca}{2} H $\&$ K to assess whether its  weakness compared to \ion{Na}{1} D is physically plausible, and a significant discrepancy  would call our \ion{Na}{1} D-based analysis into question.

We modeled the \ion{Ca}{2} lines using the same procedure described in Section \ref{sec:voigt}, but with the appropriate oscillator strengths and damping constants. We used $\{\Gamma_{\rm H}, f_{\rm osc,H},\lambda_{\rm H}\} = \{1.414 \times 10^8 \ \text{s}^{-1}, 0.3145, 3969.591~{\rm \AA}\}$ for the \ion{Ca}{2}~H line and $\{\Gamma_{\rm K}, f_{\rm osc,K},\lambda_{\rm K}\} = \{1.456 \times 10^8 \ \text{s}^{-1}, 0.6346, 3934.777~{\rm \AA}\}$ for the \ion{Ca}{2}~K line from \citet{mortonI}.

We find that the \ion{Na}{1}, \ion{Ca}{2}, and stellar velocities in the stacked spectrum are all consistent: $V_{\rm Na~I} = -22 \pm 6 ~\rm km~s^{-1}$, $V_{\rm Ca~II} = -47 \pm 15 ~\rm km~s^{-1}$, and $V_{\rm star} = -27 \pm 7~\rm km~s^{-1}$.\footnote{The stellar velocity of the stacked spectrum is not at the systemic redshift ($V_{\rm star} = 0$) because part of the galaxy is not multiply imaged.} The column density $\log(N_{\rm Ca~II}/{\rm cm}^{-2}) = 12.86 ^{+0.25} _{-0.14}$ is lower than that of $\log(N_ {\rm Na~I}/{\rm cm}^{-2}) = 13.16 ^{+0.20} _{-0.13}$, indicating
\begin{equation}
 \log \frac{N_{\rm Na~I}}{N_{\rm Ca~II}} = 0.30^{+0.24}_{-0.28}. 
\end{equation}

The \ion{Na}{1}/\ion{Ca}{2} column density ratio is a widely used diagnostic of the ISM. In Section~\ref{sec:discuss_ca}, we discuss the consequences of our measured abundance ratio, and show that its value is plausible and lies within the previously reported range for quiescent galaxies.

\begin{figure*}[t]
    \centering
    \includegraphics[width=7in]{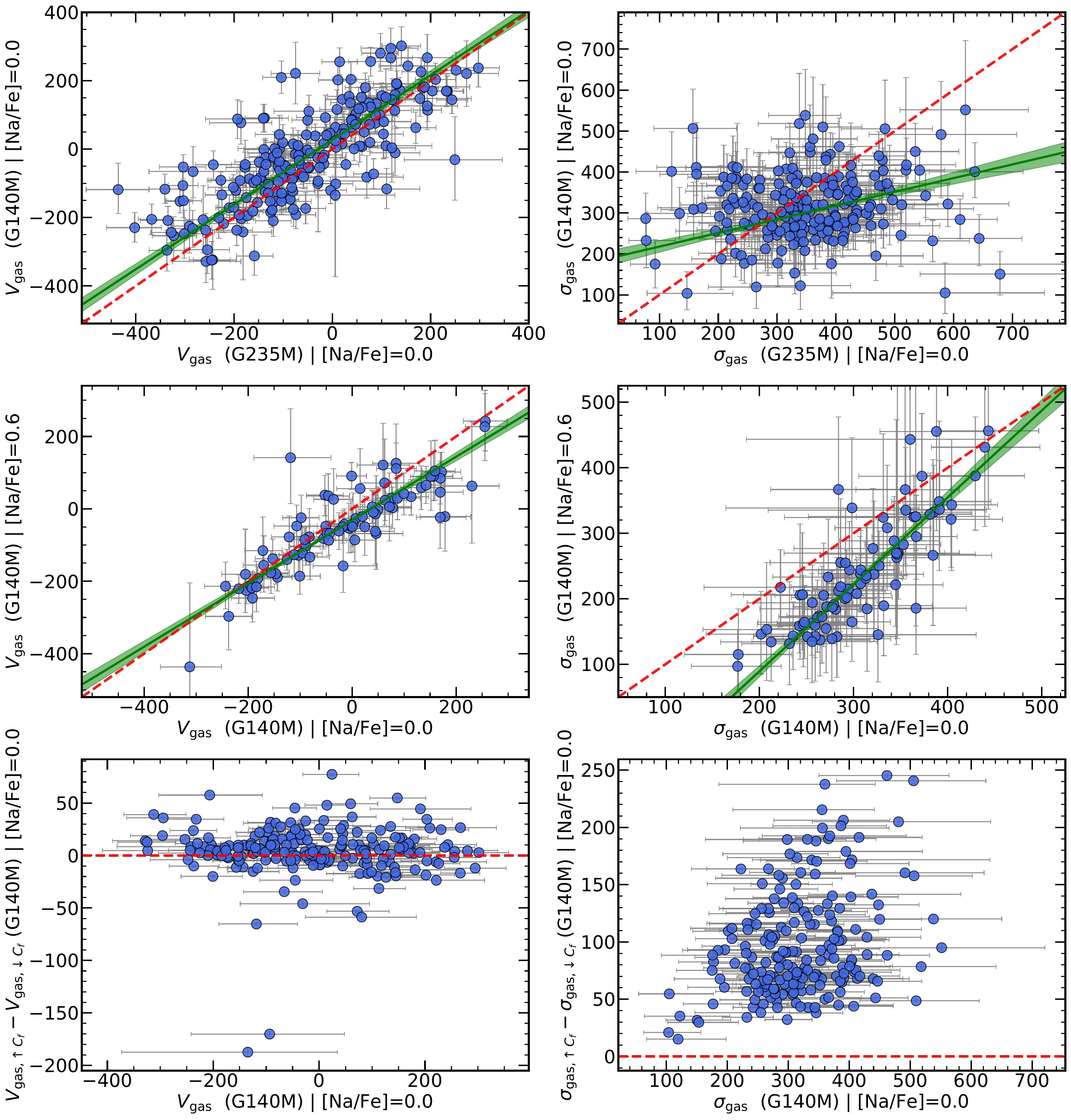}
    \caption{Tests of the robustness of the gas kinematics measured from \ion{Na}{1}~D. The left and right columns focus on the velocities ($V_{\rm gas}$) and velocity dispersions ($\sigma_{\rm gas}$), respectively. \textbf{Top row:} Comparisons of the gas kinematics extracted from the two observations (G140M and G235M), assuming ${\rm [Na/Fe]^*} = 0$. \textbf{Middle row:} Comparisons of the gas kinematics inferred when assuming [Na/Fe]* = 0 or +0.6 in the analysis, using the G140M observations. \textbf{Bottom row:} Illustration of the influence of the covering fraction $C_f$ on the gas kinematics.  We calculated the median velocity and velocity dispersion of the posterior samples when the $C_f$ value is in top 10th percentile ($\rm V_{gas,\uparrow C_f}, ~\sigma_{gas,\uparrow C_f }$) and bottom 10th percentile ($\rm V_{gas, \downarrow C_f },~\sigma_{gas,\downarrow C_f }$). In the top two rows, the red dashed line is the 1:1 line, while the green line and $1\sigma$ error band shows a linear fit. In all the plots, the units of velocity and velocity dispersion are $\rm km~s^{-1}$.}
    \label{fig:robust_}
\end{figure*}

\subsection{Robustness of Derived Kinematics}
\label{sec:gaskin}
In this section, we investigate the robustness of the \ion{Na}{1}~D kinematic measurements, $V_{\rm gas}$ and $\sigma_{\rm gas}$. Gas absorption must be detected above some threshold of significance in order for the derived kinematics to be robust and usefully precise. We selected bins for kinematic analysis using the $\Delta$AIC parameter introduced in Section~\ref{sec:namodels}. To determine an appropriate threshold, we compared the $V_{\rm gas}$ values independently derived from the G140M and G235M observations.

Specifically, we examined the relationship between $\rm min(\Delta AIC_{G140M}, \Delta AIC_{G235M})$ and $|V_{\rm gas,G140M} - V_{\rm gas,G235M}|$ to define an appropriate significance threshold, since the velocity discrepancy between the two gratings is expected to be governed by the weaker of the two detections. This analysis was performed for both assumed sodium abundances. From this comparison we adopted a threshold of $\rm min(\Delta AIC_{G140M},\Delta AIC_{G235M})=16$ and spatial bins falling below this value were excluded because they exhibit large velocity differences between the two gratings. Applying this criterion removes 15 of the 219 bins ($7\%$) from both the G140M and G235M data sets when assuming [Na/Fe]* = 0, and 135 bins ($62\%$) for [Na/Fe]* = +0.6 . We further exclude one additional bin that shows an unusually large velocity discrepancy between the two gratings.

With the suitable spatial bins defined, we turn to the reliability of the gas kinematics by exploring potential sources of systematic uncertainty. We begin by comparing the results obtained using the two different gratings to evaluate the consistency of the measured kinematics. The velocities $V_{\rm gas}$ derived from these independent data sets are well correlated (Figure~\ref{fig:robust_}, upper-left panel). Although we observe some outliers, overall the velocity measurements appear reasonably robust. In contrast, the velocity dispersions $\sigma_{\rm gas}$ do not correlate well (Figure~\ref{fig:robust_}, upper-right panel) and show significant systematic differences. We were not able to identify the cause of these differences. Spectral resolution may play a role, as it differs by a factor of 1.7 between the two gratings at the wavelength of \ion{Na}{1}~D. The spectral ``wiggles'' discussed in Section~\ref{sec:obs} could also be responsible. Although \citet{newman_bh} found that the stellar velocity dispersions derived from the full spectrum were robust to the wiggle removal, this is not necessarily true when measuring an individual spectral line.

Next, we consider the sensitivity of the gas kinematics to the assumed stellar abundance pattern ([Na/Fe]* $=0$ versus [Na/Fe]* $=+0.6$; Figure \ref{fig:robust_}, middle row). We found some systematic differences in $V_{\rm gas}$, as indicated by the sub-unity slope of the best-fit line. The velocity dispersions also show systematic differences, particularly at the low--$\sigma_{\rm gas}$ end, that reach $\approx 50-100$~km~s${}^{-1}$.

Finally, we examined the sensitivity of $\sigma_{\text{gas}}$ to the covering fraction $C_f$. Our measurements marginalize over $C_f$, but because $C_f$ is poorly constrained, the true values of $\sigma_{\rm gas}$ could differ systematically from centers of the marginalized posteriors. In the lower-right panel of Figure \ref{fig:robust_}, we plot the difference in the median $\rm \sigma_{gas}$ when restricting to posterior samples in the highest and lowest 10$\%$ of the $C_f$ posterior. This reveals a sensitivity of $\sigma_{\rm gas}$ to $C_f$ of up to $\approx 150$~km~s$^{-1}$. This behavior reflects the fact that, when $C_f$ has the minimum possible value to achieve the observed absorption depth, the absorption must be saturated, which changes the line profile. This degeneracy is also supported by the corner plots (Figure \ref{fig:corner_}), where $C_f$ and $\sigma_{\rm gas}$ exhibit covariance. We carried out an analogous test for $V_{\rm gas}$, shown in the lower-left panel of Figure~\ref{fig:robust_}, and found no systematic dependence on $C_f$. This is expected, because changes to a symmetric line profile do not affect the odd moments, such as velocity.

It is clear that the neutral gas is very turbulent, with the average velocity dispersion of the gas ranging from $\sigma_{\rm gas} \approx 330\rm~to~385~ km~s^{-1}$ among all the datasets with different assumed [Na/Fe]*. (See discussion in Section~\ref{sec:discussion} on the role of rotation-induced broadening.) However, the value of $\sigma_{\rm gas}$ in an individual spatial bin can vary by $> 100$~km~s$^{-1}$ depending on the data set and model assumptions. Such a systematic uncertainty is too large to permit detailed a kinematic interpretation. Therefore we do not consider the $\sigma_{\rm gas}$ measurements for the remainder of the paper.

Instead we will focus on the spatial distribution and velocity field of the neutral gas. Since $V_{\rm gas}$ is consistent between the two data sets, we will use an inverse-variance weighted mean of the G140M and G235M measurements, excluding measurements below the $\Delta$AIC threshold. We will show results for both the ${\rm [Na/Fe]^*} = 0$ and +0.6 models and note this distinction as needed.

\section{The Spatially Resolved Distributions of Gas, Dust, and Stars}
\label{sec:gas_dust}
\begin{figure*}[t]
    \centering
    \includegraphics[width=7in]{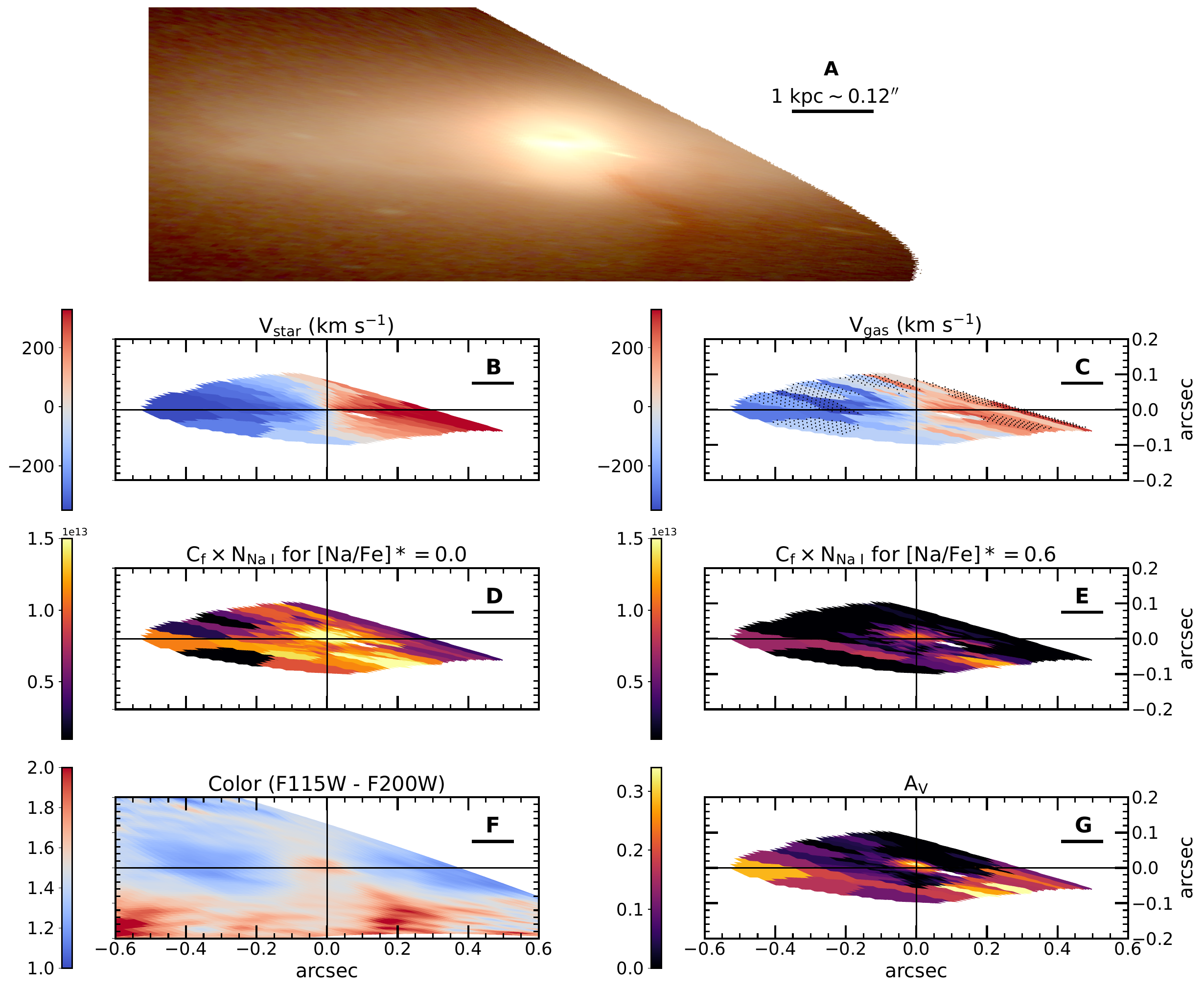}
    \caption{Maps of the distribution and kinematics of gas and dust. \textbf{(A)} Reconstructed image of the galaxy in the source plane using the NIRCam F115W, F150W, and F356W filters as in \citet{newman_bh}.  \textbf{(B)} The stellar velocity field. \textbf{(C)} The velocity of \ion{Na}{1}~D, assuming ${\rm [Na/Fe]^*} =0$. \textbf{(D)} The $C_f \times N_{\rm Na~I}$ map, assuming ${\rm [Na/Fe]^*} = 0$. This product represents the average column density over the spatial bin. \textbf{(E)} The $C_f \times N_{\rm Na~I}$ map assuming ${\rm [Na/Fe]^*} = +0.6$. \textbf{(F)} The F115W--F200W color. \textbf{(G)} The $A_V$ map derived from our SPS modeling. Dotted bins mark regions that did not meet the $\rm \Delta AIC$ threshold for robust gas kinematics. The white cross to the lower-right of the center was excluded due to contamination from SN Encore \citep{Pierel24}. The origin of the coordinate system in all spatial maps is set at the galaxy’s center, as determined from the surface brightness peak reported by \citet{newman_bh}. All gas quantities represent the average of the G140M and G235M measurements. The axis units are arcsec ($1'' = 8.6$~kpc).}
    \label{fig:maps}
\end{figure*}

Combining the results of our stellar continuum modeling and \ion{Na}{1}~D analysis, Figure~\ref{fig:maps} offers a remarkably high-resolution view of the distribution and kinematics of the gas and dust in a high-redshift quiescent galaxy. 

\subsection{Spatial Distribution}
The top panel of Figure~\ref{fig:maps} shows the starlight of MRG-M0138. This source reconstruction clearly shows a multi-component morphology: a prominent inclined disk and a compact, rounder bulge component. Because only part of the galaxy is multiply imaged at this position, the reconstruction appears diagonally truncated in the upper-right corner.

We present maps of the neutral gas kinematics in Panels B and C of Figure~\ref{fig:maps}. We map the neutral gas distribution using the product $C_f \times N$ derived from our \ion{Na}{1}~D modeling. Panels D and E show models that assume ${\rm [Na/Fe]^*} = 0$ and +0.6 for the stars, respectively. Additionally, we trace the dust using maps of the F115W--F200W color (panel F, 0.39--0.68~$\mu$m in the rest frame) and the dust attenuation $A_V$ derived from the SPS modeling (panel G). The \ion{Na}{1}~D absorption is generally stronger in regions with a redder color and higher attenuation, reinforcing the connection between the neutral gas traced by \ion{Na}{1}~D and dust.

Importantly, both the gas and dust distributions show asymmetries. The gas absorption is strongest in the bulge-dominated region, as shown by the radial plot in Figure~\ref{fig:ew_radial}, but Figure~\ref{fig:maps} shows that the peak \ion{Na}{1}~D absorption and the peak $A_V$ actually occurs off-center. The off-center dust attenuation was noted by \citet{newman_bh} who showed that this is robust to the definition of the galaxy center. 

\begin{figure*}[t]
    \centering
    \includegraphics[width=7in]{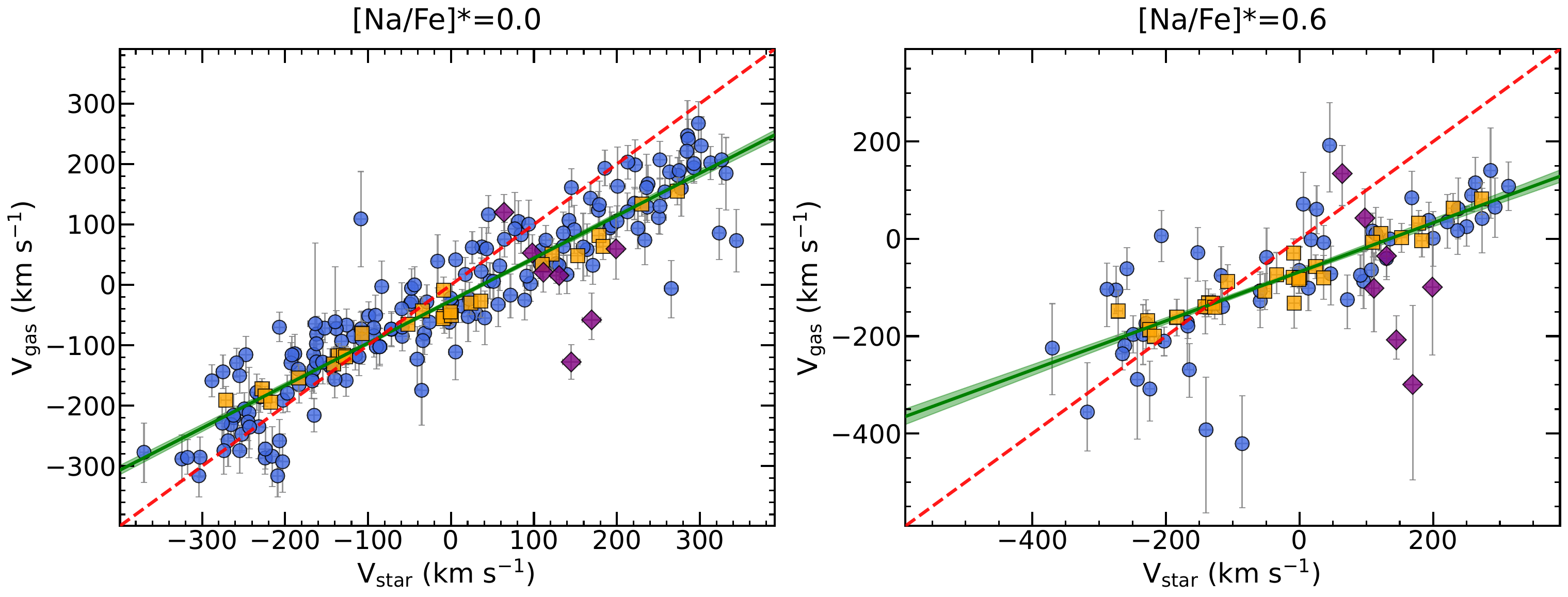}
    \caption{A comparison of the gas and stellar velocity fields. The left and right panels show results assuming ${\rm [Na/Fe]^*} = 0$ and +0.6, respectively, in the analysis. The stellar and gas velocities are clearly correlated, with the gas lagging the stellar rotation speed by an amount that depends on the assumed [Na/Fe]*. Each spatial bin that passes our significance threshold (Section~\ref{sec:namodels}) is plotted, using the average of the G140M and G235M data sets. The red dashed lines represent $V_{\rm star} = V_{\rm gas}$, and the green solid lines with $1\sigma$ error bands show linear fits. The purple diamonds indicate the region along the dust lane, while the orange squares mark the gas close to the galaxy center. These regions of enhanced gas and dust do not appear to be kinematically irregular.}
    \label{fig:starsvsgas}
\end{figure*}

At larger radii, one particularly striking feature is a well-defined dust lane visible in the color image (top panel of Figure~\ref{fig:maps}), which extends over $\rm \sim2.6~kpc$ if deprojected, assuming that it lies in the disk plane. This feature is also apparent in the color, $A_V$, and $C_f \times N$ maps, indicating a localized enhancement in gas and dust. While the narrow dust lane is most prominent, we see enhanced \ion{Na}{1}~D absorption and $A_V$ behind the dust lane that extends over $\sim120^{\circ}$ in azimuth. There may also be a feature located diametrically opposite from the dust lane, but we see a hint of it only in the \ion{Na}{1}~D map and not in the image, color, or $A_V$ maps. 

The non-axisymmetric distribution of gas and dust likely points to a recent minor merger or accretion event that deposited cold gas into the system. We will discuss this interpretion further in Section~\ref{sec:discussion}. The asymmetries in the gas distribution are robust to the assumed ${\rm [Na/Fe]^*}$. Its value mainly determines whether the asymmetries contain nearly all of the gas in the outer galaxy beyond about 2~kpc (if ${\rm [Na/Fe]^*} = +0.6$), or whether they coexist with a smooth component (if ${\rm [Na/Fe]^*} = 0$).

The presence of a smooth gas component is not necessarily inconsistent with the lack of dust attenuation observed in much of the outer galaxy. We expect \ion{Na}{1}~D absorption to be more sensitive tracer at low gas column densities. In the Milky Way, $\rm N_{H}/A_{V} \approx (2.08 \pm 0.02) \times 10^{21}~cm^{-2}~mag^{-1}$ \citep{zhu17}. We therefore expect a minimally low $A_{\rm V} \lesssim 0.1$~mag where $N_{\rm H} \lesssim 2 \times 10^{20}$~cm$^{-2}$. Using the conversion provided by \citet{moretti25}, this corresponds to $\rm N_{Na~I} \lesssim 6 \times 10^{12}$~cm$^{-2}$. Thus the lower values in the $C_f \times N$ map (Figure~\ref{fig:maps}, panel D) are expected to correspond to negligible attenuation.

\subsection{Gas Kinematics}
From the {\tt pPXF} fits, we mapped the stellar velocity $V_{\rm star}$ (Figure~\ref{fig:maps}, panel B), revealing a clear rotation pattern. The \ion{Na}{1}~D velocity field (Figure \ref{fig:maps}, panel C) also exhibits a rotation pattern, and the kinematic axis is closely aligned with that of the stars. This indicates that the neutral gas is \emph{in situ} ISM, at least mainly. Nuclear outflows, as have been found in some younger quiescent galaxies at a similar redshift \citep{davies24,belli24_, liboni25,park24}, are expected to escape perpendicular to the disk \citep{bae18,concas19,avery22,sun24}. This would result in velocity field with a kinematic axis orthogonal to the stars, which is clearly not observed in MRG-M0138.

Figure \ref{fig:starsvsgas} further illustrates the relationship between the gas and stellar kinematics. Overall, the velocity of the gas is correlated with that of the stars, but its rotation speed along a given sight line is slower. The relationship between the velocities for [Na/Fe]* = 0 is characterized by a linear fit,
\begin{equation}
    {V_{\rm gas} = (0.70 \pm 0.01) \times V_{\rm star} - (26\pm 3)\rm~km~s^{-1}},
\end{equation}
while for [Na/Fe]* = +0.6,
\begin{equation}
    {V_{\rm gas} = (0.50 \pm 0.02) \times V_{\rm star} - (68\pm 5)\rm~km~s^{-1}}.
\end{equation}
Quantitatively, therefore, the velocity lag depends on the partitioning of the absorption between stars and gas and thus the assumed ${\rm [Na/Fe]^*}$. If a higher fraction of the gas is located in the bulge compared to the stars, then the gas along a given sight line would have more bulge-like kinematics than the stars. And based on our stellar dynamical models \citep{newman_bh}, this would produce lower gas velocities. Small-scale turbulence in the gas could also provide pressure support that reduces the rotation speed required to maintain equilibrium. We will discuss the kinematics further in Section~\ref{sec:discussion}. 

In addition to the rotation speed lag, the linear fits suggest a systematic zero-point offset between the stellar and gas velocities. In principle, this could indicate a modest, galaxy-wide blueshift of the gas, potentially resulting from an outflowing component that is superposed on the dominant \emph{in situ} component. However, the magnitude of the zero-point offset depends on the assumed [Na/Fe]* and is comparable to the $\sim 40$~km~s$^{-1}$ velocity uncertainty for NIRSpec observations with the medium-resolution gratings.\footnote{\url{https://jwst-docs.stsci.edu/jwst-calibration-status/nirspec-calibration-status/nirspec-mos-calibration-status}} Consequently we refrain from drawing strong conclusions about the origin of this potential velocity offset.

In Figure~\ref{fig:starsvsgas}, we highlight the regions corresponding to the dust lane and the the central off-center clump of gas and dust. We find that these features do not display any peculiar velocities relative to the surrounding gas. Together with the global kinematic alignment of the stars and gas, this suggests that if the gas in MRG-M0138 was accreted, its angular momentum was either already nearly aligned with the galaxy disk, or it has had sufficient time to settle into corotation. We will examine the timescale for this kinematic settling and its connection to star-formation history of MRG-M0138 in Sections~\ref{sec:sed} and \ref{sec:discussion}. 

\subsection{Integrated Gas Mass Estimate}
As discussed in Section~\ref{sec:intro}, measuring the gas mass in high-redshift quiescent galaxies is difficult. \ion{Na}{1}~D has been used to infer mass outflow rates from massive, high-redshift galaxies (see Section~\ref{sec:intro}) and to draw connections to quenching, so tests of the mass estimates are important. In MRG-M0138, we have the opportunity to compare gas mass estimates derived from two methods: \ion{Na}{1}~D absorption and a prior dust continuum detection \citep{whitaker21}.

To estimate the gas mass of MRG-M0138, we use our maps of the \ion{Na}{1} column density ($N_{\rm Na~I}$) and covering fraction ($C_f$). Although $N_{\rm Na~I}$ is known to correlate linearly with the neutral hydrogen column density, $N_{\rm H} = N_{\rm H~I} + 2 N_{\rm H_2}$ \citep[e.g.,][]{Ferlet85}, the ratio depends on the physical conditions in the gas. We use the empirical conversion from \citet{moretti25}:
\begin{equation}
\log N_{\text{H I}} = \log N_{\text{Na I}} + 7.5
\end{equation}
This calibration was derived from a detailed study of an outflow from a massive galaxy at $z=2.4$ that happens to intersect the line of sight to a bright background quasar \citep{Rudie17}. The favorable geometry enabled precise measurements of the relative abundances of metals and hydrogen in the neutral phase of the outflowing gas. The outflow is measured to have a molecular hydrogen fraction of $f = 0.27$ \citep{srianand08}. Based on this, we derive and apply a slight modification of the \citet{moretti25} relation, $\log N_{\rm H} = \log N_{\rm H~I} / (1 - f) = \log N_{\rm Na~I} + 7.6$. 

We integrated the derived $C_f \times N_{\rm H}$ map in the source plane, accounting for the area subtended by each spatial bin using the lensing model. Under the assumption of a symmetric distribution, an absorbing gas parcel typically lies in front of half of the stars along its sightline. We roughly account for the star--gas geometry by multiplying the resulting absorption-based neutral hydrogen masses $M_{\rm H}$ by a factor of 2. This yields $M_{\rm H} = 4.9 ^{+3.7}_{-1.1} \times 10^7~\rm M_\odot$ for the average of the G140M and G235M data sets, assuming [Na/Fe]* = 0. Adopting models with enhanced [Na/Fe]* = +0.6 lowers the inferred neutral hydrogen mass to $M_{\rm H} \approx 1.7 \times 10^7 ~\rm M_\odot$.

We estimate the neutral gas mass fraction as
\begin{equation}
f_{\rm gas} = 1.4 \times \frac{M_{\text{H}}}{\rm M_\star},
\end{equation} 
where $M_{\text{H}}$ is the mass of atomic and molecular hydrogen, the factor 1.4 accounts for the mass of He \citep[e.g.,][]{tacconi10}, and $M_\star$ is the stellar mass of the galaxy (Section~\ref{sec:obs}). We find a gas fraction of $f_{\rm gas} = 0.031^{+0.023}_{-0.008}\%$ assuming [Na/Fe]*= 0.

We compare our gas mass fraction to that presented by \citet{whitaker21} based on a detection of MRG-M0138 at 1.3~mm, on the Rayleigh--Jeans tail of the dust emission. By assuming a molecular gas-to-dust mass ratio $\delta_{\rm GDR} = 100$, they reported a molecular gas fraction of $f_{\text{H}_2} = M_{\rm H_2} / \rm M_{\star} = 0.6 \pm 0.1 \%$. This is  $\gtrsim 12 \times$ higher than our Na-based estimate of $f_{\rm gas}$, where the lower limit indicates that we have used the $1\sigma$ upper limit on $f_{\rm gas}$. A greater discrepancy would be obtained if we were to assume [Na/Fe]* = +0.6 or add any \ion{H}{1} to the $f_{\rm H_2}$ estimate. Additionally, \citet{gobat22} reanalyzed these data and confirmed the low dust content of MRG-M0138. 

\begin{figure*}[t]
    \centering
    \includegraphics[width=7in]{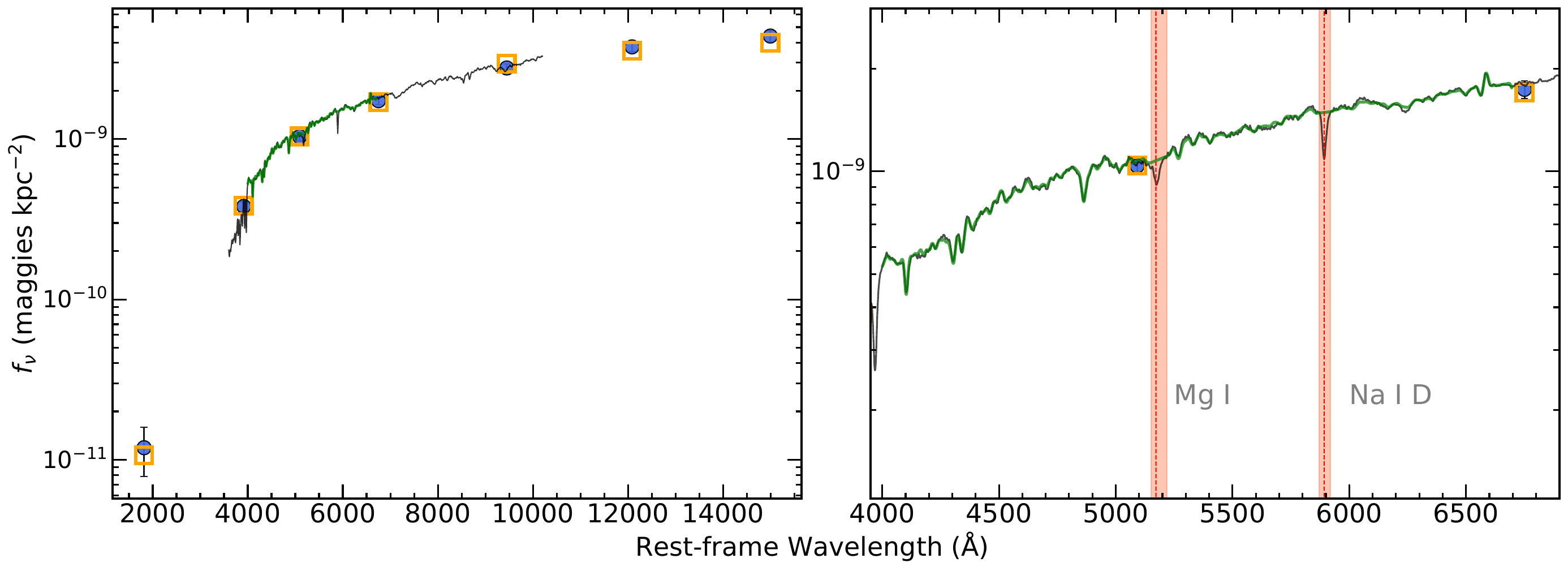}
    \caption{An example {\tt Prospector} fit for the bulge spatial bin of MRG-M0138. \textbf{Left panel:} displays the photometric measurements (blue circles) and the best-fit model (yellow squares), together with the observed spectrum (gray) and its best-fit model (green). \textbf{Right panel:} presents the same data on a reduced scale. The red dashed lines mark the specific absorption lines. The orange shaded regions indicate the wavelength intervals masked and excluded from the {\tt Prospector} fit.}
    \label{fig:prospector}
\end{figure*}

The Na- and dust-based $f_{\rm gas}$ measurements are thus significantly different. Both approaches carry substantial uncertainties. The \ion{Na}{1}~D doublet becomes optically thick at relatively low column densities \citep{schwartz04}, and once the line is saturated, increasing the column density no longer increases the absorption depth, which is then mainly set by the covering fraction \citep{heckman00, martin05, rupke05,chen10}. In addition, in realistic ISM conditions, \ion{Na}{1}~D absorption originates from multiple narrow velocity components that may be unresolved \citep{savage91, welty94, welty96}. When blended, these components can contain saturated substructure that appear as a shallower, broader feature, leading to underestimated column densities. This bias is expected to be more severe in systems with large gas velocity dispersions, like MRG-M0138. Finally, absorption measurements cannot detect any gas within highly obscured clouds. Although the galaxy is dust-poor, \ion{Na}{1} requires dust shielding to survive and so may exist in small, dusty clumps. For these reasons, gas masses derived from \ion{Na}{1}~D absorption should be regarded as lower limits. 

On the other hand, gas mass estimates derived from the dust continuum are highly sensitive to assumptions about the dust temperature ($T_{\rm dust}$) and the gas-to-dust ratio ($\delta_{\rm GDR}$). Reported values of $T_{\rm dust}$ span a broad range \citep{smith12}, with several studies finding temperatures significantly lower than the commonly adopted value of 25 K \citep{gobat18, magdis21, cochrane22}. In addition, the relation between gas and dust mass in quiescent galaxies remains poorly constrained, with observational estimates spanning a wide range of $\delta_{\rm GDR} \approx 40-1200$ \citep{whitaker21_1, morishita22, lorenzon25, spilker25}. Reconciling the Na- and dust-based $f_{\rm gas}$ estimates by varying $T_{\rm dust}$ or $\delta_{\rm GDR}$ alone would require extreme values ($T_{\rm dust} \gtrsim 100$ K or $\delta_{\rm GDR} \lesssim 15$), which are unlikely for a quiescent galaxy \citep{whitaker21_1, lorenzon25}. This instead suggests that the discrepancy arises at least in part from systematic uncertainties in the Na-based estimates.

In summary, the available tracers consistently point to an extremely low gas fraction in MRG-M0138, although the exact value remains uncertain due to systematic limitations inherent to \ion{Na}{1}~D absorption and dust continuum measurements. \ion{Na}{1}~D is a sensitive tracer of neutral gas and is effective at mapping its spatial distribution and kinematics. Our comparison of gas mass tracers suggests that Na-based masses are likely lower limits.

\section{Star-Formation Histories}
\label{sec:sed}
Motivated by signs that the some gas in MRG-M0138 may have been recently accreted, in this section we investigate the galaxy's star-formation history (SFH) using \texttt{Prospector}, a Bayesian Spectral Energy Distribution (SED) fitting code \citep{prosp1,prosp2}.   

\subsection{Modeling the Galaxy SED}

Our {\tt Prospector} fits employed the Flexible Stellar Population Synthesis (\texttt{FSPS}; \citealt{conroy09,conroy10}) models based on the MESA Isochrones and Stellar Tracks (MIST; \citealt{dotter16,choi16}), the MILES spectral library \citep{barroso11}, and a Kroupa IMF \citep{kroupa01}. The SED model includes 18 free parameters describing the contributions of stars, gas, and dust. The stellar component is characterized by the redshift, stellar mass formed, stellar velocity dispersion, metallicity (with abundance ratios assumed to be solar), and a SFH.

\begin{figure*}[t]
    \centering
    \includegraphics[width=7in]{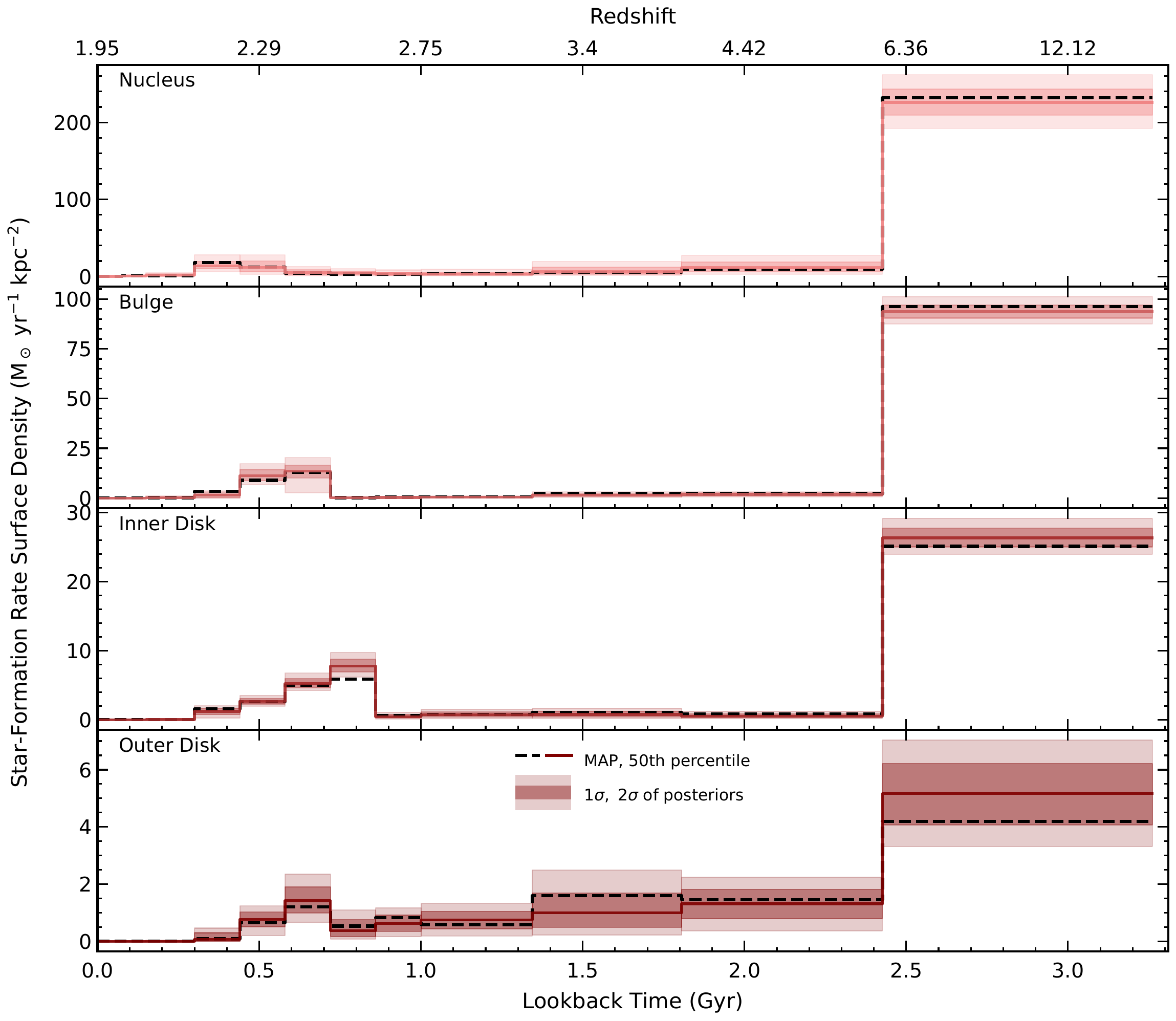}
    \caption{Non-parametric SFHs are shown for four distinct spatial regions within MRG-M0138. In each panel, the dark and light shaded regions indicate the 1$\sigma$ and 2$\sigma$ posterior intervals, respectively. The colored curves denotes the posterior medians, while the dashed black lines shows the maximum a posteriori (MAP) SFH.}
    \label{fig:sfr}
\end{figure*}

We adopted a non-parametric approach \citep{leja17,leja19}, dividing the SFH into discrete time bins without imposing any predetermined functional form. While this method increases the number of free parameters in the fit, it offers greater flexibility in capturing the true diversity of SFHs. We adopted 12 age bins: two narrow bins at young ages ($0-75$ Myr and $75-150$ Myr), followed by five linearly spaced bins between 300 Myr and 1 Gyr to better resolve recent and intermediate star formation. At older ages, we used logarithmically spaced bins extending up to $97\%$ of the age of the universe at the galaxy’s redshift. 

{\tt Prospector} fits ratios of the star formation rates (SFRs) in adjacent time bins ($\Delta \log \text{SFR}$). A Student-t ``continuity'' prior is applied to each of these ratios to encourage smooth transitions between bins while allowing for sharp changes where necessary. We used the parameters $\sigma=0.3$ and $\nu=2.0$ following \citet{leja19}. To assess the robustness of our inferred SFHs, we additionally tested a ``bursty'' prior, which is a modified version of the continuity prior with $\sigma = 1.0$. This prior does not explicitly favor sharp transitions but permits more flexible changes in star formation activity compared to the baseline continuity prior \citep{tacchella22,tacchella23}. The results do not change much with the change in prior.

Dust attenuation was modeled using a single-component \citep{calzetti00} attenuation curve applied uniformly to all stars. Therefore, only the dust parameter describing attenuation of the older stellar light is included, while the parameter for additional attenuation of young stars is fixed to zero. This is appropriate for MRG-M0138, which is an old, quiescent system with no significant young stellar population. We also experimented with dust attenuation curves that allow a variable UV slope \citep{Kriek13}, but we found this did not affect our inferred SFHs. Additionally, {\tt Prospector} can model emission lines using Gaussian profiles and analytically marginalizing their amplitudes, assuming a shared velocity offset and velocity dispersion for all lines that we considered: H$\alpha$, H$\beta$, H$\gamma$, H$\delta$, [O II] $\lambda \lambda$ 3727, 3730, [O III] $\lambda \lambda$ 4960, 5008, [N II] $\lambda \lambda$ 6550, 6585, and [S II] $\lambda \lambda$ 6718, 6733. We also included a ``jitter'' parameter that scales the reported spectroscopic flux uncertainties \citep{prosp1}. In practice, the inferred jitter values are generally close to unity, as expected given that the uncertainties were already rescaled based on our {\tt pPXF} fits. Although robust likelihood methods with mixture models \citep{prosp1} can mitigate sensitivity to outlier spectral pixels, we did not employ such an approach because outliers were already removed at the spaxel level prior to the {\tt Prospector} analysis. 

We simultaneously fit the NIRSpec spectra and matched seven-band photometry in several spatial bins described below. We fit the spectra over the rest-frame wavelength range 4000--6700~\AA~following \citet{park24}. This range covers many age-sensitive spectral features such as Balmer absorption lines. We used a polynomial order of 6 to modulate the continuum of the spectrum. We masked the \ion{Na}{1}~D and \ion{Ca}{2} K $\&$ H absorption lines, which are affected by gas absorption, as well as the \ion{Mg}{1}~b feature, which is strongly affected by $\alpha$ enhancement. To explore the posterior distribution, we use the \texttt{Dynesty} dynamic nested sampling algorithm \citep{speagle20}. 

\subsection{Spatially Resolved Star Formation Histories}
To investigate how the SFH varies across the galaxy, we measured the SFH in a series of four spatial bins. These bins were defined to probe  distinct structural components of the galaxy: the nucleus (the region within $R_{\rm e,bulge}/4$), bulge (within $R_{\rm e,bulge}$), inner disk (within $R_{\rm e,disk}$), and outer disk (within $2 R_{\rm e,disk}$).\footnote{Each region is defined as elliptical aperture with an axial ratio appropriate to the bulge or disk \citep{newman_bh}.} Each bin excludes the regions interior to it, and thus the bins are non-overlapping and the spectrophotometric data are independent. Figure~\ref{fig:prospector} presents the spectroscopic and photometric data for the nucleus spatial bin of MRG-M0138, along with the corresponding \texttt{Prospector} fit.

Figure~\ref{fig:sfr} shows that the resulting SFHs are quite uniform across the galaxy. Moreover, the bulk of MRG-M0138 is quite ancient, with most of the star-formation occurring at lookback times $> 2.5$~Gyr that correspond to $z \gtrsim 6$. Approximately 500~Myr ago, we see evidence of a rejuvenation of star formation in every spatial bin. The fraction of the stellar mass formed in this event is small, ranging from about $2\% - 9\%$ across the bins. We found that switching from the continuity prior to the bursty prior did not appreciably alter the resulting SFHs, nor did modifying the age bin structure. Therefore the qualitative features appear robust.

We quantify the significance of recent star formation enhancements using the posterior distribution of $\rm \Delta SFR_{burst}$, a metric defined to capture the strength of the burst. Specifically, $\rm \Delta SFR_{burst}$ is defined as the difference between the SFR at the peak of the secondary burst and the SFR at a lookback time of $\sim1~\rm Gyr$. The statistical significance is evaluated from the posterior probability that $\rm \Delta SFR_{burst} > 0$. This yields evidence of rejuvenation across different regions, with a significance of $\rm 2.7\sigma,~6.9\sigma,~7.1\sigma~and~ 1.7\sigma$ in the nucleus, bulge, inner disk, and outer disk, respectively. 

\citet{akhshik23} analyzed the spatially resolved SFH of MRG-M0138 based on HST grism spectroscopy. The early SFR peak that we observe is consistent with their findings. Although their analysis shows a more gradual decline in SFR and does not exhibit a secondary peak, these differences may arise from variations in methodology (e.g., the adopted SFH priors) as well as from the data itself, given that the JWST observations provide a broader wavelength coverage and much higher spectral resolution.

Since rejuvenation of quiescent galaxies is thought to be triggered by gas accretion \citep{yi05, kaviraj09, marino09}, these SFHs provide additional evidence for gas accretion and constraints on its timing. An alternative scenario is that these $\rm \sim500~Myr$-old stars formed in a younger satellite that was accreted through a minor merger. A dry merger contributing a pre-existing stellar population could , in principle, explain the secondary ``burst'' that we see in the SFH. However, as we will discuss in Section~\ref{sec:discussion}, the kinematics and distribution of the gas, together with the timing of the rejuvenation episode, favor a scenario in which externally acquired gas fueled renewed \emph{in situ} star formation. 

\section{The Environment of MRG-M0138}
\label{sec:companion}
\begin{figure}[h]
  \centering
  \includegraphics[width=\columnwidth]{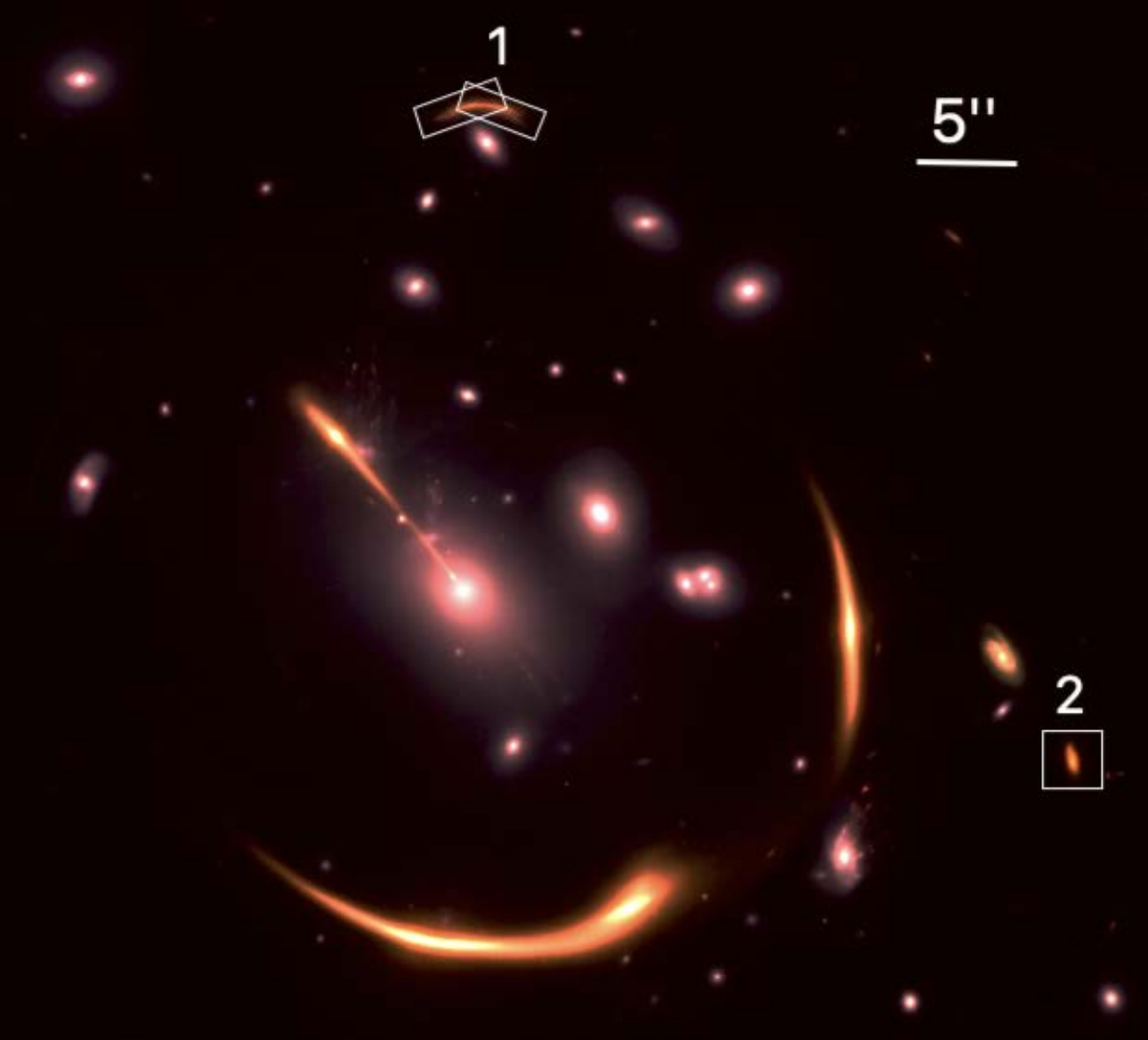}
  \caption{Two galaxies possibly associated with MRG-M0138 are identified in a JWST/NIRCam image through the F150W, F356W, and F444W filters, displayed with a logarithmic stretch. \textbf{Galaxy 1:} A multiply imaged galaxy to the north with a tentative spectroscopic redshift and a projected separation from MRG-M0138 of 77~kpc in the source plane \citep{suyu25}. \textbf{Galaxy 2:} A dusty, extremely red galaxy to the west with a photometric redshift consistent with MRG-M0138 and a projected separation of 64~kpc.}
  \label{fig:companion}
\end{figure}

Studies of $z \gtrsim 1$ quiescent and post-starburst galaxies in which CO was detected have frequently found evidence for interactions, mergers, or rejuvenation episodes \citep{belli21, suess25, zanella25}. In this context, and motivated by evidence of recent external gas accretion in MRG–M0138, we searched for nearby galaxies that could have recently interacted with it. The SFH derived from the integrated spectrophotometric data of MRG-M0138 indicates that approximately $ \sim10^{10}~\rm M_\odot~(\sim 5\%~of~M_{\star}$; Section~\ref{sec:obs}) of mass was formed during the rejuvenation episode $\rm \sim500~Myrs$ ago. If this episode is attributed to an interaction or merger, the other galaxy would need to have contributed at least this mass of gas and/or stars. Hence, we do not expect the rejuvenation to be associated with a low-mass ($\rm \lesssim 10^9~M_\odot$) companion. Our search revealed two compelling candidate companions (Figure \ref{fig:companion}).

The first candidate (labelled 1 in Figure \ref{fig:companion}), was identified in \citet{suyu25} as the ``silver lensed system.'' It consists of an arc that \citet{suyu25} identified as being formed by two blended background sources that are strongly lensed. They reported a likely spectroscopic redshift of $ z_{\rm silver}=1.945$, consistent with the redshift of MRG-M0138 at $z = 1.948$ \citep{newman_bh}. The projected separation between this source and MRG-M0138 is 77~kpc in the source plane. Because the photometry is blended and it is not entirely clear that there are two separate sources, we measured aperture photometry over the whole arc (see Figure~\ref{fig:companion}) and used {\tt Prospector} to model the data. The same set of parameters and age bins as for MRG-M0138 were adopted, but with the redshift fixed to that of MRG-M0138. Since our photometry combines three merging images, we use the sum of the magnifications $\mu \approx 32$ from the {\tt GLAFIC} lens model discussed by \citet{suyu25}.\footnote{We used this lens model because it was one of two in the set of seven lens models discussed by \citet{suyu25} to use the silver system as a constraint.} The best-fit model yields $A_{V} = 0.24$, a demagnified stellar mass of $\rm M_{*} = 9 \times 10^{9}~\rm M_\odot$, and a star-formation rate in the most recent age bin of $\rm {SFR} \approx 0.8~\rm M_{\odot}~{yr}^{-1}$, indicating that this galaxy has a moderate mass, a low SFR, and low dust attenuation.

We searched for additional companions using photometric redshift estimates generated by applying the {\tt EAZY} code \citep{Brammer08} to our 7-band HST and JWST photometry. We considered sources within $40''$ of the cluster center with a photometic redshift estimate $|z_{\rm phot} - 1.95| < 0.25$. We further restricted to galaxies with an estimated stellar mass $\rm M_* > 10^{10}$~M$_{\odot}$ to reduce contamination by the large population of low-mass galaxies, most of which will be in the foreground or background, and to focus on galaxies massive enough to have triggered the observed rejuvenation.

This search identified one additional source (labeled 2 in Figure \ref{fig:companion}), an extremely red galaxy located west of the cluster core. Its estimated photometric redshift $z_{\rm phot} = 1.82 \pm 0.18$ is consistent with MRG-M0138, it lies at a projected separation of 64~kpc in the source plane, and its magnification is $\mu = 4.5^{+1.3}_{-0.7}$ according to the \citet{ertl25} lens model. We modeled the data using \texttt{Prospector}, adopting the same configuration as for target 1. The best-fit model indicates a massive ($\rm M_*=5 \times 10^{10}~M_\odot$, approximately one-quarter that of MRG–M0138) and dusty ($A_{\rm V} = 1.9$) galaxy. The SFR in the most recent bin is relatively low ($\approx 10~{\rm M_\odot~yr}^{-1}$), but we regard this as uncertain given the high attenuation. 

Although secure spectroscopic redshifts are needed to make a unambiguous connection, these galaxies are likely associated with MRG-M0138. Given the surface density of galaxies matching the same $z_{\rm phot}$ and $\rm M_*$ criteria in the COSMOS-Web catalog \citep{shuntov25}, we would expect only 0.03 foreground or background contaminants within 77~kpc of MRG-M0138. Although the galaxies are probably not strongly interacting with MRG-M0138 at the observed epoch, they are close enough to have interacted strongly in the past, in particular 500~Myr ago when then rejuvenation event occurred. These considerations suggest that we have found galaxies capable of having acted as a gas donor through past tidal interactions.

\section{Discussion}
\label{sec:discussion}
We present high-resolution maps of the gas and dust in a quiescent galaxy at $z=1.95$, providing a detailed view of its cold ISM. Previous ALMA studies have shown that quiescent galaxies at these redshifts are typically dust-poor and, by implication, gas-poor \citep[e.g.,][]{rachel19,williams21,whitaker21}, although the relationship between gas and dust mass is not necessarily tight \citep{morishita22,lorenzon25,spilker25}. MRG-M0138 is notable in this context, as it was one of only two quiescent galaxies detected at 1.3 mm in the \citet{whitaker21} sample and was inferred to have a low gas fraction, as confirmed by \citet{gobat22}. However, prior observations lacked the spatial resolution required to resolve the dust distribution, and dust continuum observations do not enable kinematic measures.

Spatially resolved maps of the ISM provide critical insight into the origin of the remaining gas and dust in quiescent systems. In particular, they allow us to assess whether this gas represents residual ISM that survived quenching, stellar mass loss from evolved populations, or gas accreted externally. Distinguishing among these scenarios has direct implications for the completeness and long-term maintenance of quenching. It is also closely connected to the fueling of AGN feedback, which is increasingly thought to be common in high-redshift quiescent galaxies and potentially central to both the quenching process and the suppression of subsequent star formation \citep{croton06,schawinski07,somerville08, fabian12, bell12, terrazas17}. 

\subsection{Gas Kinematics}
In MRG-M0138, we detect gas absorption from the nuclear regions to the outermost radii that we probe in the disk. The gas is kinematically aligned with the stars, yet it rotates more slowly (Figures~\ref{fig:maps} and \ref{fig:starsvsgas}). Slower gas rotation relative to the circular velocity is often interpreted as the result of turbulent pressure support, which acts as a perturbation to the rotational speed in a kinematically cold disk \citep[e.g.,][]{Dalcanton10}. Turbulence could be very important in MRG-M0138, since the observed gas velocity dispersions are extremely large, $\sigma_{\rm gas} \gtrsim 200$~km~s$^{-1}$ in all spatial bins and reaching $\approx 400$~km~s$^{-1}$ or higher (Figure~\ref{fig:robust_}). However, the magnitude of any rotational lag is difficult to quantify, as it depends on the uncertain [Na/Fe]* abundance (Figure~\ref{fig:starsvsgas}), and the gas velocity dispersion is also challenging to measure (Section~\ref{sec:gaskin}).

Moreover, the observed line width $\sigma_{\rm gas}$ is likely to be strongly affected by rotational broadening, a consequence of projection through a highly inclined galaxy, PSF convolution, and spatial binning. In the fiducial stellar dynamical model of \citet{newman_bh}, when these effects are taken into account, the smallest stellar velocity dispersion expected to be measured is $\sigma_* = 189$~km~s${}^{-1}$, which is much higher than the intrinsic stellar velocity dispersion of about 70~km~s$^{-1}$ in the disk-dominated region. All of these effects make it difficult to accurately measure the gas turbulence and quantitatively interpret the rotation lag. The most robust conclusion is that the rotation fields of the gas and stars are kinematically aligned.

This corotation of the gas and stars disfavors the presence of large-scale gas outflows in MRG-M0138, as have been reported in many massive galaxies at $z\gtrsim2$ \citep{eugenio24, belli24, davies24, wu25, valentino25, liboni25}. We found that the systemic velocity of the gas may be slightly blueshifted with respect to the stars, but the magnitude of any such offset is comparable to the systematic uncertainties (Section~\ref{sec:gaskin}). A handful of spatial bins show gas that may be blueshifted relative to the local stellar velocity, with differences of up to $\sim \rm 260~km~s^{-1}$ (Figure \ref{fig:starsvsgas}, left panel), but the gas kinematics are uncertain in these bins: they fail our significance threshold under the sodium-enhanced scenario, [Na/Fe]* = +0.6.  We therefore do not interpret these as outflows. We also see no sign of coherent inflows, including in the dust lane, which would be detectable as regions with $V_{\rm gas} > V_{\rm stars}$ in Figure~\ref{fig:starsvsgas}. The gas may be slowly infalling, but its kinematics are dominated by disk motion.

\subsection{Spatial Distribution of Gas and Dust}
Spatially, the gas in MRG-M0138 does not appear to be in a equilibrium configuration, due to the presence of a prominent one-sided dust lane, a trailing wedge-shaped structure in the outer disk, and an off-nuclear concentration of gas and dust. This argues against stellar mass loss as the dominant origin of the gas, since material shed by stars would be expected to follow the stellar spatial distribution more closely. While the stars must have released substantial amounts of gas and dust since quenching, most of this material is likely heated and thus not observable in a cold phase, as long established for local early-type galaxies \citep{vanDokkum95}. Disturbed gas morphologies and dust features in local early-type galaxies are commonly interpreted as signatures of recent accretion or merger events \citep{sage93, serra12, kreckel12, sancisi08}. 

Secondary episodes of star formation, or rejuvenation, are also often attributed to mergers or gas accretion \citep{marino09, kaviraj09, paudel23, wang25}. The star formation histories that we inferred with \texttt{Prospector} reveal a minor, galaxy-wide rejuvenation episode approximately 500 Myr ago (Figure~\ref{fig:sfr}), providing a plausible timing for at least one gas accretion event. The time since accretion must be long enough for the gas to have been torqued by the stellar disk into corotation, as we observe. This timescale has been explored in simulations by \citet{vandeVoort15}. Once the gas supply is largely shut off, they find that kinematic settling occurs on a timescale of roughly $6t_{\rm dyn}$, and potentially twice as long to reach complete alignment. Defining the dynamical time as $t_{\rm dyn}=2\pi R/v_{\rm circ}$, we estimate $t_{\rm dyn}\rm \approx50~Myr$ in the outer spatial bins. This implies a kinematic settling timescale of $\sim 300 - 600$ Myr, strikingly close to the epoch of minor rejuvenation. This suggests that the gas has had enough time to kinematically settle. If the gas arrived on a nearly corotating orbit, still less time may have been required.

The spatial distribution of the gas may take longer to reach equilibrium. If the gas in MRG-M0138 was accreted over an extended period, for instance, through the slow fallback of tidal gas ejected during a merger or interaction, the continued infall of material could delay relaxation to an equilibrium configuration \citep{vandeVoort15,davis16}. Indeed, we identify two galaxies that may have interacted with MRG-M0138 (Section~\ref{sec:companion}). The nearer galaxy is remarkably red, dusty, and presumably gas-rich. While the projected separation may not support a strong ongoing interaction, the galaxies may have been closer when the accretion occurred $\sim 500$~Myr ago. Although further observations and simulations are needed to explore this scenario in detail, interactions between these galaxies and MRG-M0138 are certainly a plausible origin of the accreted gas. As highlighted by \citet{whitaker25}, globular cluster systems serve as an independent probe of merger and accretion histories, and a similar analysis is currently underway for MRG-M0138 (Matthews et al., in prep).

Despite the overall disturbed morphology of the gas and dust, one of its main components is a remarkably clear dust lane that is readily visible in the NIRCam imaging (Figure \ref{fig:maps}A).  Studies of local ETGs show that the net dust depletion time can extend to a $\sim$Gyr \citep{Martini13}, implying that dust acquired $\sim500$~Myr ago could plausibly still be observable. More surprising would be for the dust to have persisted in this coherent, linear configuration for roughly ten dynamical times \citep{davis16}. Although the dust lane appears sharply defined in the image, the $N_{\rm Na~I}$ and color maps (Figure \ref{fig:maps}, panel D-F) are more sensitive to lower densities. Together with the velocity field, these show that the dust lane is the leading edge of an extended wedge of gas and dust in the outer disk, suggesting that it may be a dynamical structure formed by interactions between accreted gas and pre-existing ISM. Alternatively, gas accretion may be episodic, and the material in the dust lane could have been acquired more recently than the rejuvenation event.

\subsection{Implications for Feedback and Quenching}

Our results reinforce the interpretation of \citet{whitaker21} that quenching at high redshifts is highly effective at removing or heating the interstellar medium. In particular, our analysis suggests that at least one of the dust detections in that work arises from material accreted after quenching, rather than from residual gas left over from the star-forming phase. Moreover, while the decline of the gas or dust reservoirs of quenched galaxies is often modeled as a smooth gas consumption history \citep[e.g.,][]{gobat18}, MRG-M0138 is a reminder that the external accretion of gas can complicate this picture. Its little remaining gas is probably not related to the main quenching event.

The gas may, however, be related to the maintenance of quiescence. The off-nuclear clump of gas and dust is located extremely close to the galaxy center, at a projected distance of only $\approx 0.01'' \approx 90$~pc. Given its proximity, this material could soon feed the $\rm 6 \times 10^9~M_\odot$ supermassive black hole at the center of MRG-M0138 \citep{newman_bh}, which is currently inactive. Such accretion could plausibly trigger an episode of black hole feedback. Low-luminosity AGN are common in local early-type galaxies, often referred to as ``red geysers'' \citep{cheung16,roy18,ilha22,frank23}, where episodic feedback events may provide sufficient energy to prevent gas cooling and star formation. In the prototype red geyser system, Akira \citep{cheung16}, the gas fueling the central black hole appears to have been supplied via a tidal interaction. Evidence for episodic AGN feedback in high-redshift quiescent galaxies has also been discussed by \citet{taylor26}, who propose that outflows may later fall back as inflows if the gas cannot escape completely, thereby reigniting AGN activity on timescales of $\sim$40 Myr.

These neutral gas outflows appear to be more common and more powerful in galaxies that were quenched within the last $\sim1$~Gyr \citep{park24, taylor24, sun24, taylor26}. Our SFH analysis indicates that the main quenching event in MRG–M0138 occurred $\gtrsim 2.5$~Gyr ago at $z \gtrsim 6$ (Figure~\ref{fig:sfr}), placing it among the earliest galaxies to quench. The old age of MRG-M0138 could explain its lack of outflows. Furthermore, these population studies imply that the gas in MRG-M0138 has probably not been continually recycled since the main quenching period, and they therefore support our interpretation that the gas was acquired afterward.

\subsection{The Weak Calcium Absorption in MRG-M0138}
\label{sec:discuss_ca}
While \ion{Na}{1} gas absorption is widespread across MRG-M0138, we detect much weaker \ion{Ca}{2} H and K absorption only through stacking (Section~\ref{sec:ca}). We estimate a lower column density of \ion{Ca}{2}, with $\log (N_{\rm Na~I}/N_{\rm Ca~II}) = 0.19^{+0.22}_{-0.32}$. This is near the upper limit of the range $-0.8 \lesssim \log (N_{\rm Na~I}/N_{\rm Ca~II}) \lesssim 0.3$ recently observed in a sample of $z\sim2$ galaxies, both quiescent and star-forming \citep{liboni25}. Therefore the much weaker \ion{Ca}{2} H $\&$ K absorption compared to \ion{Na}{1}~D is not highly unusual.

The abundance of calcium in interstellar gas is highly sensitive to the ionization state and dust depletion, whereas sodium is comparatively less affected \citep{hobbs75, bertin93, savage96, welty99, kondo06, loon09}. Specifically, cold ($T \sim 30$ K), dense ($n_H \geq 10^3~{\rm cm}^{-3}$) diffuse molecular clouds, where most of the gas-phase calcium is depleted onto grains, typically exhibit \ion{Na}{1}/\ion{Ca}{2} $\geq100$, whereas values two or more orders of magnitude lower are characteristic of warmer ($T\sim10^3$ K), lower-density environments where much of the calcium remains in the gas phase \citep{hobbs76, bertin93, crawford92}. In addition, shocks can liberate calcium from grain surfaces \citep{routly52, siluk74, crawford89, vallerga93, sembach94}. The relatively high $N_{\rm Na~I}/N_{\rm Ca~II}$ that we measure in MRG-M0138 compared to other quiescent galaxies at similar redshift \citep{liboni25} suggests cooler conditions. Qualitatively, the old stellar population (Figure~\ref{fig:sfr}) and quiescent black hole \citep{newman_bh} in MRG-M0138 are consistent with a weaker far-ultraviolet radiation field and less photoelectric heating. Moreover, the gas absorption in about half of the \citet{liboni25} galaxies is blueshifted, indicating that it is produced in an outflow where the physical conditions may be more favorable for dust grain destruction (hence less \ion{Ca}{2} depletion).

\section{Conclusions and Summary}
\label{sec:conclusion}
We present the first highly resolved picture of gas and dust in a $z \sim 2$ quiescent galaxy. We analyzed JWST/NIRSpec IFU observations of the remarkable lensed galaxy MRG-M0138, complemented by HST and JWST/NIRCam imaging in seven filters. The high magnification of this galaxy ($\mu \sim 29$ at center) enables us to extract high-quality spectra in 219 spatial bins. We traced the spatial distribution and kinematics of the neutral gas using \ion{Na}{1}~D absorption, as well as the dust distribution by mapping the attenuation. Our main results are summarized below.

\begin{itemize}
    \item We detected \ion{Na}{1}~D absorption (beyond the stellar contribution) tracing neutral gas in most spatial bins across a wide range of plausible stellar [Na/Fe]* values (Figure~\ref{fig:ew_radial}). In contrast, \ion{Ca}{2} H and K absorption is much weaker and only detectable in stacked spectra. 
    
    \item The \ion{Na}{1}~D absorption traces a smooth velocity field with clear rotation that is aligned with the stellar velocity field but slower, likely due to turbulent pressure support. We find no evidence for large-scale inflows or outflows like those seen in some high-redshift quiescent galaxies, possibly reflecting the galaxy’s old stellar population and early quenching. 

    \item The gas and dust show a highly inhomogeneous distribution, including a one-sided radial dust lane, a trailing wedge of clumpy material in the disk, and a clump near but offset from the nucleus (Figure~\ref{fig:maps}). This disturbed morphology favors an external accretion origin for most or all of the gas and dust, rather than residual material from quenching or stellar mass loss.
    
    \item Using \texttt{Prospector}, we modeled the spatially resolved SFH of MRG–M0138, finding rapid quenching $\sim2.5$ Gyr ago followed by $\sim2$ Gyr of quiescence. A minor rejuvenation episode $\sim500$ Myr ago is detected, possibly triggered by external gas accretion. This timescale matches that required for accreted gas to be torqued into corotation with the stellar disk. 

    \item We identified two galaxies at projected separations of 61–77 kpc from MRG–M0138, one with a tentative spectroscopic redshift and one with a consistent photometric redshift, making them plausible sources of tidally acquired gas. 

    \item MRG–M0138 was one of only two $z\sim2$ quiescent galaxies detected in ALMA dust continuum emission by \citet{whitaker21}. Our spatially resolved analysis suggests that this gas is not a remnant of the early quenching phase, but instead was likely accreted recently from a nearby companion. This reinforces the conclusion by \citet{whitaker21} that quenching in massive high-redshift galaxies efficiently consumed or removed gas.

    \item The off-nuclear gas clump lies $\sim0.1$~arcsec ($\approx90$ pc) from the central SMBH, making it a plausible fuel source for future AGN feedback. This highlights a pathway in which external gas accretion triggers AGN activity. Recurrent feedback may help maintain quiescence in MRG–M0138, akin to local ``red geysers,'' and AGN feedback seeded by mergers or accretion may similarly power some of the winds seen in recently quenched $z\sim2$ galaxies.

\end{itemize}

 A comprehensive analysis of the full INQUEST-JWST sample is currently underway. By extending the approaches developed in this study to a larger sample of lensed, quiescent galaxies, we will assess the incidence and properties of inflows and outflows, as well as the origin of any residual interstellar gas. Ultimately we aim to test whether hidden gas accretion and feedback cycles represent a ubiquitous mechanism for maintaining quiescence at high redshift.

\begin{acknowledgments}
We thank Masamune Oguri, Yoshinobu Fudamoto, Nicholas Foo, Brenda Frye, and Shun Nishida for sharing the magnification of the silver system from their lens model. We thank Sherry Suyu for sharing the magnification and source plane position of the red galaxy discussed in Section 6. This work is based on observations made with the NASA/ESA/CSA James Webb Space Telescope. The data were obtained from the Mikulski Archive for Space Telescopes at the Space Telescope Science Institute, which is operated by the Association of Universities for Research in Astronomy, Inc., under NASA contract NAS 5-03127 for JWST. These observations are associated with programs \#2345 and \#6549. Support for programs \#2345 and \#6549 was provided by NASA through a grant from the Space Telescope Science Institute, which is operated by the Association of Universities for Research in Astronomy, Inc., under NASA contract NAS 5-03127.
\end{acknowledgments}

\software{\texttt{AstroPy} \citep{astropy13,astropy18,astropy22}, \texttt{SciPy} \citep{scipy20}, \texttt{emcee} \citep{emcee13}, \texttt{dynesty} \citep{speagle20}, \texttt{pPXF} \citep{ppxf23}, \texttt{Prospector} \citep{prosp1,prosp2}}

\bibliographystyle{aasjournalv7}
\bibliography{main}{}

\end{document}